\documentclass[a4paper,12pt]{article}
\usepackage{mathrsfs}
\usepackage{amsfonts}
\usepackage{overcite}
\usepackage{amsmath,amssymb,enumerate,epsfig,bbm}
\usepackage[margin=30pt]{caption}

\allowdisplaybreaks[4]
\begin{document}

\title{\bf Indecomposable representations and oscillator realizations of the exceptional Lie algebra $G_2$ }
\author{{ Hua-Jun Huang, You-Ning Li, and Dong Ruan }{\footnote{dongruan@tsinghua.edu.cn}}   \\
\\
{\it Department of Physics, Tsinghua University, Beijing 100084, P.R. China}
\\}
\maketitle

\vspace{1cm}

\begin{abstract}
In this paper various representations of the exceptional Lie algebra $G_2$ are investigated in a purely algebraic manner, and multi-boson/multi-fermion realizations are obtained. Matrix elements of the master representation, which is defined on the space of the universal enveloping algebra of $G_2$, are explicitly determined. From this master representation, different indecomposable representations defined on invariant subspaces or quotient spaces with respect to these invariant subspaces are discussed. Especially, the elementary representations of $G_2$ are investigated in detail, and the corresponding six-boson realization is given. After obtaining explicit forms of all twelve extremal vectors of the elementary representation with the highest weight $\mathbf{\Lambda}$, all representations with their respective highest weights related to $\mathbf{\Lambda}$ are systematically discussed. For one of these representations the corresponding five-boson realization is constructed. Moreover, a new three-fermion realization from the fundamental representation (0,1) of $G_2$ is constructed also.
\end{abstract}

\vfill
\newpage 

\section{INTRODUCTION}\label{sec:Intro}
The exceptional Lie algebra $G_2$ has many physical applications. The first successful application is in atomic spectroscopy, where Racah\cite{Racah} used $G_2$ to classify the states of atomic $f^n$ configurations according to the Lie algebra chain $so(7)$ $\supset$ $G_2$ $\supset$ $so(3)$ and to simplify the calculations of the matrix elements of the Coulomb interaction. Soon afterwards, Racah's idea and method was directly applied to nuclear physics \cite{Jahn} to classify the orbital states of nuclear $f^n$ configurations in Russell-Saunders coupling according to the Lie algebra chain $u(7)$ $\supset$ $so(7)$ $\supset$ $G_2$ $\supset$ $so(3)$.\cite{Flowers,RD4} Morrison {\it et al.}\cite{IBM} discussed the interacting boson model including $s, d, g, i$ bosons by use of $u(28)$ $\supset$ $u(7)$ $\supset$ $so(7)$ $\supset$ $G_2$ $\supset$ $so(3)$. In particle physics, $G_2$ symmetry is drawing more attention recently and is the topic of even more works\cite{Dall,Carlos,Ferre,Yang-Mills,Csaki}. These applications motivate us to study $G_2$.



Considerable research has been done investigating $G_2$. Irreducible matrix representations of $G_2$ have been obtained by embedding $G_2$ into $A_{6}$\cite{A6}, and in the $G_2\supset su(3)$ basis\cite{Rowe1} as well as the $G_2\supset so(4)$ basis,\cite{Rowe2} respectively. The Casimir operators of degrees 2 and 6 of $G_2$ with its generators written in terms of the subalgebra $A_{2}$ were obtained.\cite{Casimir} In recent years, some works on $G_2$ have been put forward such as new types of representations with solitary and edge-minimal bases,\cite{Edge} twist quantization,\cite{twistQ} and exceptional quantum subgroups.\cite{Quantum} However, much less is known about indecomposable representations of $G_2$. The indecomposable representations (i.e. reducible but not completely reducible representations) have been found useful in physics for a long time.\cite{Niederer,Chacon,Flato1,Flato2,HZSun,Lesimple,Alkalaev} Many authors have investigated the indecomposable representations of various algebras: Lorentz algebra, \cite{Zhelobenko,Gelfand} Euclidean algebra,\cite{Gruber-Henne,Douglas} Poincar\'e algebra,\cite{Lenczewski} Lie superalgebras,\cite{HZSun,Fu} quantum groups,\cite{Chari,Xiao} the Diamond Lie algebra,\cite{diamond} and so on. Especially, Gruber and co-workers\cite{Gruber0,Gruber1,Gruber2,Gruber3} have studied in detail the indecomposable representations of $A_{1}$, $A_{2}$, and $D_{2}$ on the spaces of their respective universal enveloping algebras and on the quotient spaces with respect to the invariant subspaces. Recently the indecomposable representations appeared to be topical in the representation theory of the nonlinear Lie algebras.\cite{RD1,RD2} The indecomposable representations can also be used to construct various boson realizations of the (nonlinear) Lie algebras.\cite{Gruber6,CPSUN,RD3}

In this paper we will first investigate the master representation of $G_2$ on the space of its universal enveloping algebra, and various representations which are defined on different invariant subspaces and on quotient spaces with respect to different ideals. Then from these representations we will construct different oscillator realizations in terms of six bosons, five bosons, and three fermions, respectively.

This paper is arranged as follows: In Sec. \ref{sec:genepro}, the general procedure of obtaining inhomogeneous boson realizations (IHBRs)\cite{RD2} from the master representation or from the induced representations is briefly reviewed, which is applicable to any Lie algebra. In Sec. \ref{sec:master}, explicit forms of the matrix elements of the master representation $\rho$ of $G_2$ are determined, and its various indecomposable representations and the corresponding IHBRs are discussed respectively. In Sec. \ref{sec:elementary}, the matrix elements of the elementary representation $d_{\mathbf{\Lambda}}$ with $\mathbf{\Lambda}$ being the highest weight, as an induced representation of $\rho$, are calculated. Then a six-boson realization from $d_{\mathbf{\Lambda}}$ is constructed. In Sec. \ref{sec:extremal}, all twelve extremal vectors of $d_{\mathbf{\Lambda}}$ are determined. They may define various invariant subspaces which carry various elementary sub-representations. Then we investigate two series of representations on different quotient spaces. Different cases of each series are discussed, a five-boson realization is obtained from one of them. Moveover, to find realizations with fewer boson or fermion operators, we manage to construct a new three-fermion realization from the fundamental representation (0,1) as an induced representation of $d_{\mathbf{\Lambda}}$. A summary and some discussions are given in the final section.

In this paper, $\mathbb{N}$ denotes the set of positive integers, $\mathbb{Z}^{+}$ the set of non-negative integers, and $\mathbb{C}$ the set of complex numbers.

\section{THE GENERAL PROCEDURE}\label{sec:genepro}
Consider a Lie algebra $\mathfrak{g}$ with generators $T_i$ $(i=1,2,\ldots,m,m=$ dim $\mathfrak{g})$. According to the Poincare-Birkhoff-Witt theorem, a basis for the universal enveloping algebra $\mathfrak{U}(\mathfrak{g})$ of $\mathfrak{g}$ can be naturally chosen as the following set of ordered elements:
\begin{equation}\label{eq:bas1}
\{X(i_1,i_2,\ldots,i_m)=T_{1}^{i_1}T_{2}^{i_2}\dots T_{m}^{i_m}|i_1,i_2,\ldots,i_m\in\mathbb{Z}^{+}\}.
\end{equation}
We call $\{X(i_1,i_2,\ldots,i_m)\}$ the PBW basis of $\mathfrak{g}$. The left action of the generators $T_a$ upon this PBW basis (\ref{eq:bas1}) gives
\begin{equation}\label{eq:act}
    \rho(T_a)X(i_1,i_2,\ldots,i_m)=T_{a}T_{1}^{i_1}T_{2}^{i_2}\dots T_{m}^{i_m}=\sum_{i_{1}',\ldots,i_{m}'}{\rho(T_a)_{i_1,i_2,\ldots,i_m}^{i_{1}',i_{2}',\ldots,i_{m}'}X(i_{1}',i_{2}',\ldots,i_{m}')}.
\end{equation}
Here we obtain a representation $\rho$ of $\mathfrak{g}$ on $\mathfrak{U}(\mathfrak{g})$, called the master representation. The matrix elements $\rho(T_{a})_{i_{1}i_{2}\ldots i_{m}}^{i_{1}'i_{2}'\ldots i_{m}'}$ can be explicitly determined by the commutation relations of $\mathfrak{g}$. Let ${I}$ be a left ideal of $\mathfrak{U}(\mathfrak{g})$. On the quotient space $\mathfrak{U}(\mathfrak{g})/{I}$, $\rho$ may induce a representation, and different choices of $I$ enable us to obtain various representations of $\mathfrak{g}$.

Now we can construct the IHBR of $\mathfrak{g}$ from the master representation $\rho$ on $\mathfrak{U}(\mathfrak{g})$ or from the representations on $\mathfrak{U}(\mathfrak{g})/{I}$. Since the matrix elements $\rho(T_a)_{i_1,i_2,\ldots,i_m}^{i_{1}',i_{2}',\ldots,i_{m}'}$ in Eq. ($\ref{eq:act}$) are related to $m$ independent parameters $i_1,i_2,\dots,i_m$, we need $m$ pairs of independent boson operators $\{a_{i}^{\dag},a_{i}\}\;(i=1,2,\ldots,m)$ to define a boson Fock space $\mathcal{F}_{b}$ with basis
\begin{equation}\label{eq:Fbas}
\mathcal{F}_{b}:\{|i_{1},i_{2},\ldots,i_{m}\rangle \equiv {a_{1}^{\dag}}^{i_1}{a_{2}^{\dag}}^{i_2}\dots{a_{m}^{\dag}}^{i_m}|0\rangle\;| i_{1},i_{2},\ldots,i_{m}\in \mathbb{Z}^{+}\},
\end{equation}
where $|0\rangle$ is the vacuum state of $\mathcal{F}_{b}$, and $a_{i}|0\rangle = 0$. These boson operators satisfy the usual boson commutation relations:
\begin{eqnarray}\label{eq:bosoncomm}
&[a_{i},a_{j}^{\dag}]=\delta_{ij},\quad
[a_{i},a_{j}]=[a_{i}^{\dag},a_{j}^{\dag}]=0,\\
&\hat{n}_{i}\equiv a_{i}^{\dag}a_{i}, \quad [\hat{n}_{i}, a_{j} ]= - \delta_{ij} a_{j}, \quad [\hat{n}_{i}, a_{j}^{\dag}]= \delta_{ij} a_{j}^{\dag},
\end{eqnarray}
with $\hat{n}_{i}$ being the particle number operator of the $i$th boson.

Then the mapping $\Phi: \mathfrak{U}(\mathfrak{g})\longrightarrow\mathcal{F}_{b}$ or $\Phi: \mathfrak{U}(\mathfrak{g})/I\longrightarrow\mathcal{F}_{b}$ defined by
\begin{equation}\label{eq:map}
    \Phi(X(i_1,i_2,\ldots,i_s))= |i_{1},i_{2},\ldots,i_{s}\rangle, \quad s\leq m
\end{equation}
is an automorphism to $\mathfrak{U}(\mathfrak{g})$ or $\mathfrak{U}(\mathfrak{g})/I$. Let
\begin{equation}\label{eq:gamma}
    \Gamma(T_i)= \Phi \rho(T_i) \Phi^{-1}.
\end{equation}
It is obvious that $\Gamma(T_i)$ satisfy the same commutation relations as the generators $T_i$ of $\mathfrak{g}$, i.e.,
\begin{equation}\label{eq:commu}
    [\Gamma(T_i),\Gamma(T_j)] = \Gamma([T_i,T_j]).
\end{equation}
Therefore, Eq. ($\ref{eq:gamma}$) defines a Fock representation of $\mathfrak{g}$ on $\mathcal{F}_{b}$:
\begin{equation}\label{eq:fockrep}
    \Gamma(T_a)|i_{1},i_{2},\ldots,i_{m}\rangle =\sum_{i_{1}',\ldots,i_{m}'}{\rho(T_a)_{i_1,i_2,\ldots,i_m}^{i_{1}',i_{2}',\ldots,i_{m}'}|i_{1}',i_{2}',\ldots,i_{m}'\rangle}.
\end{equation}
Note that the matrix elements $\rho(T_a)_{i_1,i_2,\ldots,i_m}^{i_{1}',i_{2}',\ldots,i_{m}'}$ in Eq. (\ref{eq:fockrep}) are just the same as those in Eq. (\ref{eq:act}). It is easy to obtain the corresponding IHBR of $\mathfrak{g}$ with the help of the following formulas:
\begin{eqnarray}\label{eq:bosonrela}
&&a_{k}^{\dag}|\ldots,i_{k},\ldots\rangle = |\ldots,i_{k}+1,\ldots\rangle,\nonumber\\
&&a_{k}|\ldots,i_{k},\ldots\rangle  =
i_{k}|\ldots,i_{k}-1,\ldots\rangle ,\\
&&\hat{n}_{k}|\ldots,i_{k},\ldots\rangle  =
i_{k}|\ldots,i_{k},\ldots\rangle. \nonumber
\end{eqnarray}

\section{THE MASTER REPRESENTATION OF $G_2$}{\label{sec:master}}
$G_2$ is a fourteen-dimensional simple Lie algebra of rank two. \cite{Humphreys} Its root system $\Sigma$ has twelve roots $\left\{\pm\alpha_i|i=1,2,\dots,6\right\}$. Since the Cartan subalgebra of $G_2$ is two-dimensional, the six positive roots in $\Sigma$ may be chosen as:
\begin{eqnarray}
\alpha_1=(\frac{1}{4},-\frac{\sqrt{3}}{4}),\quad\alpha_2=(\frac{1}{4},-\frac{1}{4\sqrt{3}}),\quad
\alpha_3=(\frac{1}{2},0), \nonumber\\
\alpha_4=(\frac{1}{4},\frac{1}{4\sqrt{3}}),\quad
\alpha_5=(\frac{1}{4},\frac{\sqrt{3}}{4}),\quad\alpha_6=(0,\frac{1}{2\sqrt{3}}),
\end{eqnarray}
where $\alpha_1, \alpha_6$ are two simple roots.

\begin{figure}[!hbp]
\begin{center}
\includegraphics[width=0.4\textwidth]{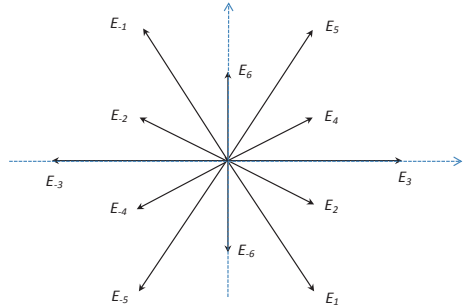}
\caption{Cartan-Weyl basis of $G_2$}\label{fig:roots}
\end{center}
\end{figure}

The Cartan-Weyl basis of $G_2$ reads
\begin{equation}
\{E_{\pm\alpha_{1}},E_{\pm\alpha_{2}},E_{\pm\alpha_{3}},E_{\pm\alpha_{4}},E_{\pm\alpha_{5}},E_{\pm\alpha_{6}},H_{1},H_{2}\}.
\end{equation}
In this paper, we will use the convenient notations $E_{\pm i} \equiv E_{\pm \alpha_{i}}$, which are shown in Fig. \ref{fig:roots}. The subset $\{H_{1},H_{2}\}$ forms the Cartan subalgebra $\mathfrak{h}$ of $G_2$.

The commutation relations of $G_2$ are
\begin{eqnarray}
\lbrack H_{i},E_{\alpha } \rbrack &=&\alpha^{(i)} E_{\alpha },\quad \textrm{with }\alpha^{(i)}\equiv \alpha(H_{i}),\;i=1,2; \label{commurela1}\\
\lbrack E_{\alpha },E_{\beta } \rbrack &=&
\left\{\begin{array}{ll}
\sum_{i}\alpha^{(i)} H_{i}, & \textrm{if}\; \alpha + \beta = 0 , \\
N_{\alpha,\beta}E_{\alpha+\beta}, & \textrm{if}\; \alpha + \beta \in \Sigma,
\end{array}\right.\label{commurela2}
\end{eqnarray}
where coefficients $N_{\alpha,\beta}$ are respectively
\begin{equation}
N_{6,1}=N_{6,4}=N_{4,2}=N_{1,5}=\frac{1}{2\sqrt{2}},\quad
N_{6,2}=\frac{1}{\sqrt{6}}
\end{equation}
with the relations $N_{\alpha,\beta}=-N_{-\alpha,-\beta}$.

The PBW basis of the universal enveloping algebra $\Omega$ of $G_2$ can be chosen as the following set of ordered elements
\begin{eqnarray}{\label{eq:standardbasisofmaster}}
\Omega :\{X(m_1,\dots,k_2)\equiv E_{-1}^{m_{1}} E_{-2}^{m_{2}} E_{-3}^{m_{3}}
E_{-4}^{m_{4}} E_{-5}^{m_{5}} E_{-6}^{m_{6}}E_{1}^{n_{1}} E_{2}^{n_{2}} E_{3}^{n_{3}}
E_{4}^{n_{4}} E_{5}^{n_{5}} E_{6}^{n_{6}}H_{1}^{k_1}H_{2}^{k_2}\nonumber\\|m_{i},n_{i},k_{1},k_{2}\in \mathbb{Z}^{+},i=1,2,\ldots,6\}.
\end{eqnarray}
In this paper we call the order from $E_{- 1}$ to $H_2$ in Eq. (\ref{eq:standardbasisofmaster})``the standard order'', and let $\mathbf{1}$ denote the identity operator, which is with all parameters simulataneously being zero, i.e., $\mathbf{1}=X(0,0,\ldots,0)$. Before further computation, it is useful to derive the following general commutation relations between any two Cartan-Weyl elements of a Lie algebra $\mathfrak{g}$:
\begin{eqnarray}
&&\lbrack H_{i}, E_{\alpha}^{n}\rbrack = n\alpha^{(i)} E_{\alpha}^{n}, \label{eq:commurela1}\\
&&\lbrack E_{\alpha }, E_{-\alpha}^{n} \rbrack = \sum_{i=1}^{2}n\alpha^{(i)}E_{-\alpha}^{n-1}H_{i}
+\frac{1}{2}n(n-1)(\alpha,-\alpha)E_{-\alpha}^{n-1},\\
&&\lbrack E_{\alpha }, E_{\beta }^{n} \rbrack = \sum_{m}C_{n}^{m}\left(\prod_{i=1}^{m}N_{\alpha+(i-1)\beta,\beta}\right)E_{\beta}^{n-m}E_{\alpha+m\beta},
\label{eq:commurela2}
\end{eqnarray}
where $n,m\in\mathbb{N}$, $m\leq n$ and $m \leq 3$, and $C_{n}^{m}\equiv \frac{n!}{m!(n-m)!}$ is the binomial coefficient.

Using above equations, the master representation $\rho$ of $G_2$ can be
directly calculated. The explicit matrix elements of $\rho$ are very complicated and are put in Appendix, where we have used $X_{(mnk)}$ as abbreviation of $X(m_{1},\dots,k_2)$, $X_{m_i+1}$ as abbreviation of $X(\dots,m_i+1,\dots)$. Notice that the matrix elements for the two generators $\{H_1,H_2\}$ of $\mathfrak{h}$ are
\begin{eqnarray}
\left\{\begin{array}{ll}
\rho(H_{1}) X_{(mnk)}
=X_{k_{1}+1} + \left(\sum_{i=1}^{6}\alpha_{i}^{(1)}(n_{i}-m_{i})\right)
X_{(mnk)},\\
\rho (H_{2}) X_{(mnk)}
=X_{k_{2}+1} + \left(\sum_{i=1}^{6}\alpha_{i}^{(2)}(n_{i}-m_{i})\right)
X_{(mnk)}.
\end{array}\right.
\label{eq:CSA}
\end{eqnarray}
Hence any element of the PBW basis $\{X_{(mnk)}\}$ of $G_2$ is not the common eigenvector of $\{H_1,H_2\}$.

By making use of Eq. (\ref{eq:bosonrela}), we can obtain directly from $\rho$ an inhomogeneous fourteen-boson realization of $G_2$. However, we are interested in realizations with fewer boson pairs. Thus, different representations on quotient spaces with respect to different left ideals need investigating.

In $\rho$ the parameters $k_1$, $k_2$, $n_5$, $n_6$, $n_{\widehat{45}}\equiv n_4 + n_5$, and $n_{\widehat{15}} \equiv n_1+n_2+2n_3+n_4+n_5$ can be increased only. This means that $\rho$ is indecomposable in any of these parameters. Therefore, each of the subspaces $V_{K_1}$, $V_{K_2}$, $V_{N_5}$, $V_{N_6}$, $V_{N_{\widehat{45}}}$, and $V_{N_{\widehat{15}}}$ of $\Omega$, with their respective bases being
\begin{eqnarray*}
\begin{array}{ll}
 V_{K_1}: &\{E_{-1}^{m_1}\ldots E_{1}^{n_1}\ldots E_{5}^{n_5} E_{6}^{n_6}H_{1}^{K_{1} + k_{1}}H_{2}^{k_2}|K_{1}\in\mathbb{N}\}, \\
 V_{K_2}: &\{E_{-1}^{m_1}\ldots E_{1}^{n_1}\ldots E_{5}^{n_5} E_{6}^{n_6}H_{1}^{k_{1}}H_{2}^{K_2 + k_2}|K_{2}\in\mathbb{N}\}, \\
 V_{N_5}: &\{E_{-1}^{m_1}\ldots E_{1}^{n_1}\ldots E_{5}^{N_5+ n_5}E_{6}^{n_6}H_{1}^{k_{1}}H_{2}^{k_2}|N_{5}\in\mathbb{N}\}, \\
 V_{N_6}: &\{E_{-1}^{m_1}\ldots E_{1}^{n_1}\ldots E_{5}^{n_5}E_{6}^{N_{6}+n_6}H_{1}^{k_{1}}H_{2}^{k_2}|N_{6}\in\mathbb{N}\}, \\
 V_{N_{\widehat{45}}}: &\{E_{-1}^{m_1}\ldots E_{1}^{n_1}\ldots E_{5}^{N_{\widehat{45}}-n_4+ n_5}E_{6}^{n_6}H_{1}^{k_{1}}H_{2}^{k_2}|N_{\widehat{45}}\in\mathbb{N}\}, \\
 V_{N_{\widehat{15}}}: &\{E_{-1}^{m_1}\ldots E_{1}^{n_1}\ldots E_{5}^{(N_{\widehat{15}}-n_1-n_2-2n_3-n_4)+n_5}E_{6}^{n_6}H_{1}^{k_{1}}H_{2}^{k_2}|N_{\widehat{15}}\in\mathbb{N}\},
\end{array}
\end{eqnarray*}
forms an invariant subspace of $\Omega$ under the action of $\rho$. These subspaces carry sub-representations induced by $\rho$. Furthermore, $\rho$ can also induce the representations on the quotient spaces of $\Omega$ with respect to these invariant subspaces, and their matrix elements can be obtained by formally setting
\begin{eqnarray*}
\begin{array}{ll}
 E_{-1}^{m_1}\ldots E_{1}^{n_1}\ldots E_{5}^{n_5} E_{6}^{n_6}H_{1}^{K_{1}}H_{2}^{k_2}=0, & \textrm{for}\; \Omega/V_{K_1},\\
 E_{-1}^{m_1}\ldots E_{1}^{n_1}\ldots E_{5}^{n_5} E_{6}^{n_6} H_{1}^{k_{1}}H_{2}^{K_2}=0, & \textrm{for}\; \Omega/V_{K_2},\\
 E_{-1}^{m_1}\ldots E_{1}^{n_1}\ldots E_{5}^{N_5}E_{6}^{n_{6}}H_{1}^{k_{1}}H_{2}^{k_2}=0, & \textrm{for} \; \Omega/V_{N_5},\\
 E_{-1}^{m_1}\ldots E_{1}^{n_1}\ldots E_{5}^{n_5}E_{6}^{N_{6}}H_{1}^{k_{1}}H_{2}^{k_2}=0, & \textrm{for} \; \Omega/V_{N_6},\\
 E_{-1}^{m_1}\ldots E_{1}^{n_1}\ldots E_{5}^{N_{\widehat{45}}-n_4}E_{6}^{n_6}H_{1}^{k_{1}}H_{2}^{k_2}=0, & \textrm{for}\; \Omega/V_{N_{\widehat{45}}},\\
 E_{-1}^{m_1}\ldots E_{1}^{n_1}\ldots E_{5}^{N_{\widehat{15}}-n_1-n_2-2n_3-n_4}E_{6}^{n_6}H_{1}^{k_{1}}H_{2}^{k_2}=0, & \textrm{for}\; \Omega/V_{N_{\widehat{15}}}
\end{array}
\end{eqnarray*}
in $\rho$, respectively. Different values of $K_1$, $K_2$, $N_5$, $N_6$, $N_{\widehat{45}}$, and $N_{\widehat{15}}$ correspond to different representations on their respective quotient spaces. For example, if we choose $K_1=1$, then $H_1$ does not appear in the PBW basis of the quotient space $\Omega/V_{{K}_1=1}$, that is, its basis involves only thirteen generators of $G_2$. The corresponding IHBR needs thirteen independent boson pairs. Moreover, since a sum of any combination of these invariant subspaces is still an invariant subspace $V$ of $\Omega$, a new representation on the quotient space $\Omega/V$ can be induced. Different choices of the sum spaces determine different representations. This makes it possible to find various indecomposable representations with their respective PBW bases containing only 13, 12, 11, 10, 9, 8, 7, or 6 generators of $G_2$ as follows:
\begin{eqnarray}\label{eq:quotients}
&&13 \textrm{ bosons}: \;\; \Omega/V_{\overline{K}_1},\;\Omega/V_{\overline{K}_2},\;\Omega/V_{\overline{N}_5},\;\Omega/V_{\overline{N}_6}; \nonumber\\&&
12 \textrm{ bosons}: \;\; \Omega/\cup_{\overline{K}_1 \overline{K}_2},\; \Omega/\cup_{\overline{K}_1 \overline{N}_5},\;
\Omega/\cup_{\overline{K}_1 \overline{N}_6},\;\Omega/\cup_{\overline{K}_2 \overline{N}_5},\;\nonumber\\&&
\qquad\quad\quad\;\;\;\;\;\;\Omega/\cup_{\overline{K}_2 \overline{N}_6},\;\Omega/\cup_{\overline{N}_5 \overline{N}_6},\;\Omega/V_{\overline{N}_{\widehat{45}}}; \nonumber\\&&
11 \textrm{ bosons}: \;\; \Omega/\cup_{\overline{K}_1 \overline{K}_2 \overline{N}_5},\; \Omega/\cup_{\overline{K}_1 \overline{K}_2 \overline{N}_6},\;
\Omega/\cup_{\overline{K}_1 \overline{N}_5 \overline{N}_6},\;\Omega/\cup_{\overline{K}_2 \overline{N}_5 \overline{N}_6},\nonumber\\&&
\qquad\quad\quad\;\;\;\;\;\; \Omega/\cup_{\overline{K}_1 \overline{N}_{\widehat{45}}},\;\Omega/\cup_{\overline{K}_2 \overline{N}_{\widehat{45}}},\;\Omega/\cup_{\overline{N}_{\widehat{45}}\overline{N}_{6}}; \nonumber\\&&
10 \textrm{ bosons}: \;\; \Omega/\cup_{\overline{K}_1 \overline{K}_2 \overline{N}_5 \overline{N}_6},\; \Omega/\cup_{\overline{K}_1 \overline{K}_2 \overline{N}_{\widehat{45}}},\;
\Omega/\cup_{\overline{K}_1 \overline{N}_{\widehat{45}} \overline{N}_6},\;\Omega/\cup_{\overline{K}_2 \overline{N}_{\widehat{45}} \overline{N}_6};\nonumber\\&&
\;\;9 \textrm{ bosons}: \;\; \Omega/\cup_{\overline{K}_1 \overline{K}_2 \overline{N}_{\widehat{45}} \overline{N}_6},\; \Omega/V_{\overline{N}_{\widehat{15}}}; \nonumber\\&&
\;\;8 \textrm{ bosons}: \;\; \Omega/\cup_{\overline{K}_1 \overline{N}_{\widehat{15}}},\;\Omega/\cup_{\overline{K}_2 \overline{N}_{\widehat{15}}},\; \Omega/\cup_{\overline{N}_{\widehat{15}} \overline{N}_6};\; \nonumber\\&&
\;\;7 \textrm{ bosons}: \;\; \Omega/\cup_{\overline{K}_1 \overline{K}_2 \overline{N}_{\widehat{15}}},\; \Omega/\cup_{\overline{K}_1 \overline{N}_{\widehat{15}} \overline{N}_6 },\; \Omega/\cup_{\overline{K}_2 \overline{N}_{\widehat{15}} \overline{N}_6};\;\nonumber\\&&
\;\;6 \textrm{ bosons}: \;\;  \Omega/ \cup_{\overline{K}_{1}\overline{K}_2 \overline{N}_{\widehat{15}} \overline{N}_6}.
\end{eqnarray}
Here for convenience we have used the short notations: $V_{\overline{K}_1}\equiv V_{K_1=1}$ and $\cup_{\overline{K}_1 \overline{K}_2 \dots}\equiv V_{K_1=1}+V_{K_2=1}+\dots$.

Just like $\rho$ in Eq. (\ref{eq:CSA}), because of the existence of $X_{k_i+1}$, the basis elements in most of these representations are not the common eigenvectors of $\left\{H_1,H_2\right\}$, except for the representations on $\Omega/\cup_{\overline{K}_1 \overline{K}_2}$, $\Omega/\cup_{\overline{K}_1 \overline{K}_2 \overline{N}_5}$, $\Omega/\cup_{\overline{K}_1 \overline{K}_2 \overline{N}_6}$, $\Omega/\cup_{\overline{K}_1 \overline{K}_2 \overline{N}_5 \overline{N}_6}$, $\Omega/\cup_{\overline{K}_1 \overline{K}_2 \overline{N}_{\widehat{45}}}$, $\Omega/\cup_{\overline{K}_1 \overline{K}_2 \overline{N}_{\widehat{45}} \overline{N}_6}$, $\Omega/\cup_{\overline{K}_1 \overline{K}_2 \overline{N}_{\widehat{15}}}$, and $\Omega/ \cup_{\overline{K}_{1}\overline{K}_2 \overline{N}_{\widehat{15}} \overline{N}_6}$.

\begin{figure}[!hbp]
\begin{center}
\includegraphics[width=0.65\textwidth]{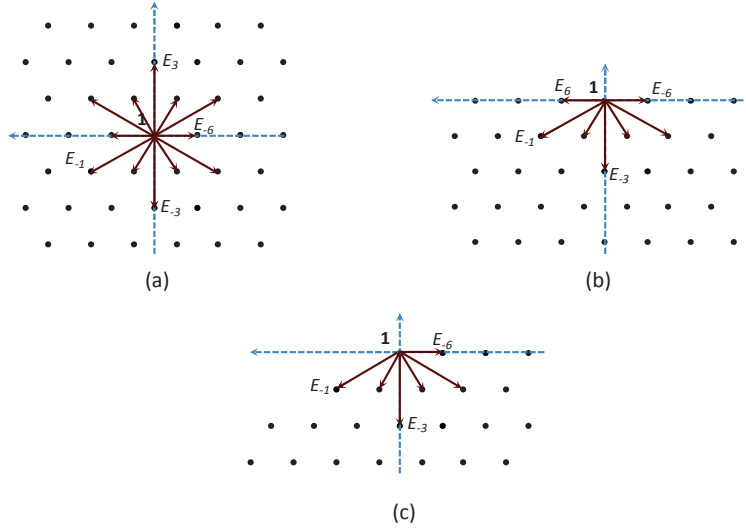}
\caption{A dot denotes a weight of the weight diagram. 
(a) The weight diagram of the representations on  $\Omega/ \cup_{\overline{K}_{1}\overline{K}_2}$, $\Omega/\cup_{\overline{K}_1 \overline{K}_2 \overline{N}_5}$, $\Omega/\cup_{\overline{K}_1 \overline{K}_2 \overline{N}_6}$, $\Omega/\cup_{\overline{K}_1 \overline{K}_2 \overline{N}_5 \overline{N}_6}$, $\Omega/\cup_{\overline{K}_1 \overline{K}_2 \overline{N}_{\widehat{45}}}$, and $\Omega/\cup_{\overline{K}_1 \overline{K}_2 \overline{N}_{\widehat{45}} \overline{N}_6}$. (b) The weight diagram of the representation on $\Omega/\cup_{\overline{K}_1 \overline{K}_2 \overline{N}_{\widehat{15}}}$. (c) The weight diagram of the elementary representation on $\Omega/\cup_{\overline{K}_1 \overline{K}_2 \overline{N}_{\widehat{15}} \overline{N}_6}$, with $\mathbf{1}$ being its highest weight vector.}{\label{fig:4diagrams}}
\end{center}
\end{figure}

In the case of $\Omega/ \cup_{\overline{K}_{1}\overline{K}_2}$ with the basis
\begin{eqnarray*}
 \{ \, \bar{X}_{(mn)}\equiv E_{-1}^{m_{1}} E_{-2}^{m_{2}} E_{-3}^{m_{3}} E_{-4}^{m_{4}} E_{-5}^{m_{5}} E_{-6}^{m_{6}}E_{1}^{n_{1}} E_{2}^{n_{2}} E_{3}^{n_{3}}
E_{4}^{n_{4}} E_{5}^{n_{5}} E_{6}^{n_{6}} \, \},
\end{eqnarray*}
Eq. (\ref{eq:CSA}) now becomes
\begin{eqnarray}\label{eq:weight}
\left\{\begin{array}{ll}
\rho(H_{1}) \bar{X}_{(mn)}
=\left(\sum_{i=1}^{6}\alpha_{i}^{(1)}(n_{i}-m_{i})\right)
\bar{X}_{(mn)},\\
\rho (H_{2})\bar{ X}_{(mn)}
=\left(\sum_{i=1}^{6}\alpha_{i}^{(2)}(n_{i}-m_{i})\right)
\bar{X}_{(mn)}.
\end{array}
\right.
\end{eqnarray}
This set of equations shows that the basis element $\bar{X}_{(mn)}$ in $\Omega/ \cup_{\overline{K}_{1}\overline{K}_2}$ is the common eigenvector of $\{H_1,H_2\}$, with their respective eigenvalues being
\begin{eqnarray}\label{eq:M}
\left\{\begin{array}{ll}
\bar{M}_{1}
&=\sum_{i=1}^{6}\alpha_{i}^{(1)}(n_{i}-m_{i})\\
&=\frac{1}{4}(n_1+n_2+2n_3+n_4+n_5-m_1-m_2-2m_3-m_4-m_5),\\
\bar{M}_{2}
&=\sum_{i=1}^{6}\alpha_{i}^{(2)}(n_{i}-m_{i})\\
&=\frac{1}{4\sqrt{3}}(-3n_1-n_2+n_4+3n_5+2n_6+3m_1+m_2-m_4-3m_5-2m_6).
\end{array}
\right.
\end{eqnarray}
The vector $\bar{\mathbf{M}} \equiv (\bar{M}_1, \bar{M}_2)$ is the weight of $\bar{X}_{(mn)}$. It is obvious that the weight of $\mathbf{1}$ is $\mathbf{0}$. 
The weight diagram of all weights $\bar{\mathbf{M}}$ of $\Omega/ \cup_{\overline{K}_{1}\overline{K}_2}$ is shown in Fig. \ref{fig:4diagrams}(a).

The similar arguments about $\Omega/ \cup_{\overline{K}_{1}\overline{K}_2}$ can also be applied to the remaining seven quotient spaces: $\Omega/\cup_{\overline{K}_1 \overline{K}_2 \overline{N}_5}$, $\Omega/\cup_{\overline{K}_1 \overline{K}_2 \overline{N}_6}$, $\Omega/\cup_{\overline{K}_1 \overline{K}_2 \overline{N}_5 \overline{N}_6}$, $\Omega/\cup_{\overline{K}_1 \overline{K}_2 \overline{N}_{\widehat{45}}}$, $\Omega/\cup_{\overline{K}_1 \overline{K}_2 \overline{N}_{\widehat{45}} \overline{N}_6}$, $\Omega/\cup_{\overline{K}_1 \overline{K}_2 \overline{N}_{\widehat{15}}}$, and $\Omega/ \cup_{\overline{K}_{1}\overline{K}_2 \overline{N}_{\widehat{15}} \overline{N}_6}$. For these cases, it is obvious that not all twelve parameters $ m_{i}, n_{i}$ $(i=1,2,\ldots,6)$ appear. The weight diagrams of $\Omega/\cup_{\overline{K}_1 \overline{K}_2 \overline{N}_5}$, $\Omega/\cup_{\overline{K}_1 \overline{K}_2 \overline{N}_6}$, $\Omega/\cup_{\overline{K}_1 \overline{K}_2 \overline{N}_5 \overline{N}_6}$, $\Omega/\cup_{\overline{K}_1 \overline{K}_2 \overline{N}_{\widehat{45}}}$, and $\Omega/\cup_{\overline{K}_1 \overline{K}_2 \overline{N}_{\widehat{45}} \overline{N}_6}$ are the same as that of $\Omega/ \cup_{\overline{K}_{1}\overline{K}_2}$. However, the weight diagrams of $\Omega/\cup_{\overline{K}_1 \overline{K}_2 \overline{N}_{\widehat{15}}}$ and $\Omega/{\cup}_{\overline{K}_1 \overline{K}_2 \overline{N}_{\widehat{15}} \overline{N}_6}$ are different from that of $\Omega/{\cup}_{\overline{K}_1 \overline{K}_2 }$, which are shown in Fig. \ref{fig:4diagrams} (b) and (c) respectively.

The weight diagram of $\Omega/\cup_{\overline{K}_1 \overline{K}_2 \overline{N}_{\widehat{15}}}$ occupies half of the full weight lattice of $G_2$, i.e., its weight has an upper bound in one direction. In the case of $\Omega/{\cup}_{\overline{K}_1 \overline{K}_2 \overline{N}_{\widehat{15}} \overline{N}_6}$, there exists a highest weight vector $\mathbf{1}$ with its weight being $\mathbf{0}$, owing to the fact that all six generators $E_{-i}$ ($i=1$, 2,..., 6) corresponding respectively to the six negative roots appear in the basis elements of
$$
 \Omega/{\cup}_{\overline{K}_1 \overline{K}_2 \overline{N}_{\widehat{15}} \overline{N}_6}:\left\{E_{-1}^{m_{1}} E_{-2}^{m_{2}} E_{-3}^{m_{3}}E_{-4}^{m_{4}} E_{-5}^{m_{5}} E_{-6}^{m_{6}}\right\}.
 $$
Actually, the representation on $\Omega/{\cup}_{\overline{K}_1 \overline{K}_2 \overline{N}_{\widehat{15}} \overline{N}_6}$ is a special case (with the highest weight being zero) of the elementary representations, which will be discussed in the next section.

\section{THE ELEMENTARY REPRESENTATIONS OF $G_2$}\label{sec:elementary}
The PBW basis (\ref{eq:standardbasisofmaster}) of $G_2$ can be factored into a product of three subspaces: ${\Omega} = \Omega_{-}{\Omega}_{+}\mathcal{H}$, with
\begin{eqnarray}\label{factorspace}
   \Omega_{-} : &  & \{X^{-}(m_1,m_2,\ldots,m_6)=E_{-1}^{m_1}E_{-2}^{m_2}\dots E_{-6}^{m_6}|m_i\in \mathbb{Z}^{+},i=1,2,\ldots,6\},
 \nonumber \\
   \Omega_{+} : &  & \{X^{+}(n_1,n_2,\ldots,n_6)=E_{1}^{n_1}E_{2}^{n_2}\dots E_{6}^{n_6}|n_i\in \mathbb{Z}^{+},i=1,2,\ldots,6\}, \nonumber \\
   \mathcal{H} : &  & \{X^{h}(k_1,k_2)=H_{1}^{k_1}H_{2}^{k_2}|k_1,k_2\in \mathbb{Z}^{+}\}.
\end{eqnarray}
If on $\Omega$ one imposes the conditions
\begin{equation}\label{impo}
\left\{\begin{array}{ll}
\rho(H_j) \mathbf{1} = \Lambda_{j}\mathbf{1},\quad \Lambda_{j}\in\mathbb{C},\; j=1,2,\\
\rho(E_i) \mathbf{1} = 0,  \quad i=1,2,\ldots,6,
\end{array}\right.
\end{equation}
then one may obtain a quotient space $Z(\mathbf{\Lambda})=\Omega/I(\mathbf{\Lambda})$, where $I(\mathbf{\Lambda})$ is the left ideal of $\Omega$ generated by $\{E_1, E_2,\ldots,E_6,H_1-\Lambda_1 \mathbf{1},H_2-\Lambda_2 \mathbf{1}\}$. Eq. (\ref{impo}) shows that $\mathbf{\Lambda}\equiv(\Lambda_1,\Lambda_2)$ is the weight of the vector $\mathbf{1}$, and $\mathbf{1}$ is the highest weight vector of $Z(\mathbf{\Lambda})$. The representation on $Z(\mathbf{\Lambda})$ is called the elementary representation\cite{Gruber1}, denoted by $d_{\mathbf{\Lambda}}$. When the highest weight $\mathbf{\Lambda}=\mathbf{0}$, the representation $d_{\mathbf{\Lambda}=\mathbf{0}}$ becomes the case of $\Omega/{\cup}_{\overline{K}_1 \overline{K}_2 \overline{N}_{\widehat{15}} \overline{N}_6}$ in the last section.

Under the conditions (\ref{impo}), the matrix elements of the master representation $\rho$ (see Appendix) reduce into those of the elementary representation $d_{\mathbf{\Lambda}}$. Here, using the abbreviations: $X_{(m)}^{-} \equiv X^{-}(m_1,m_2,\ldots,m_6)$ and $X^{-}_{m_{i}+k} \equiv X^{-}(m_1,\ldots,m_{i}+k,\ldots,m_6)$ in $\Omega_{-}$, the explicit matrix elements of $d_{\mathbf{\Lambda}}$ are as follows:
\begin{eqnarray}\label{eq:dlambda}
&&\rho (E_{-1})X^{-}_{(m)} =X^{-}_{m_{1}+1} ,\nonumber\\
&&\rho (E_{-2})X^{-}_{(m)} =X^{-}_{m_{2}+1} ,\nonumber\\
&&\rho (E_{-3})X^{-}_{(m)} =X^{-}_{m_{3}+1} ,\nonumber\\
&&\rho (E_{-4})X^{-}_{(m)} =X^{-}_{m_{4}+1} - \frac{1}{2\sqrt{2}}m_{2}
X^{-}_{m_{2}-1,m_{3}+1} ,\nonumber\\
&&\rho (E_{-5})X^{-}_{(m)} =X^{-}_{m_{5}+1} + \frac{1}{2\sqrt{2}}m_{1}
X^{-}_{m_{1}-1,m_{3}+1} ,\nonumber\\
&&\rho (E_{-6})X^{-}_{(m)} =X^{-}_{m_{6}+1} - \frac{1}{2\sqrt{2}} m_{1}  X^{-}_{m_{1}-1,m_{2}+1} - \frac{1}{\sqrt{6}}m_{2}  X^{-}_{m_{2}-1,m_{4}+1} \nonumber\\
&& \qquad\qquad +\frac{1}{8\sqrt{3}}{m_{2}(m_{2}-1)} X^{-}_{m_{2}-2,m_{3}+1} - \frac{1}{2\sqrt{2}} m_{4}  X^{-}_{m_{4}-1,m_{5}+1},\nonumber\\
&&\rho (E_{1})X^{-}_{(m)} =\frac{1}{4}m_{1}\left[\Lambda_{1}-{\sqrt{3}}\Lambda_{2}
-\frac{1}{2}(m_{1}-1+m_{2}+m_{3}-m_{5}-m_{6})\right]X^{-}_{m_{1}-1}\nonumber\\
&& \qquad\qquad +
\frac{1}{2\sqrt{2}}m_{2}X^{-}_{m_{2}-1,m_{6}+1}- \frac{1}{8\sqrt{3}}{m_{2}(m_{2}-1)} X^{-}_{m_{2}-2,m_{4}+1} \; \nonumber\\
&& \qquad\qquad +\frac{1}{48\sqrt{6}}{m_{2}(m_{2}-1)(m_{2}-2)} X^{-}_{m_{2}-3,m_{3}+1} \nonumber\\
&& \qquad\qquad - \frac{1}{8}m_{2} m_{4} X^{-}_{m_{2}-1,m_{4}-1,m_{5}+1} - \frac{1}{2\sqrt{2}}m_{3} X^{-}_{m_{3}-1,m_{5}+1 } ,\nonumber\\
&&\rho (E_{2}) X^{-}_{(m)} = - \frac{1}{4\sqrt{3}}m_{1}m_{4}X^{-}_{m_{1}-1,m_{2}+1,m_{4}-1}\;\nonumber\\
&& \qquad\qquad +
\frac{1}{16\sqrt{6}}{m_{1}m_{4}(m_{4}-1)}X^{-}_{m_{1}-1,m_{3}+1,m_{4}-2}
- \frac{1}{8}m_{1}m_{5}X^{-}_{m_{1}-1,m_{4}+1,m_{5}-1}
\nonumber\\
&& \qquad\qquad + \frac{1}{4\sqrt{6}}m_{1} m_{6}
\left(\Lambda_{2}-\frac{m_{6}-1}{4\sqrt{3}}\right)
X^{-}_{m_{1}-1,m_{6}-1}\;
\nonumber\\ && \qquad\qquad
+ \frac{1}{4}m_{2}\left[\Lambda_{1}
-\frac{1}{\sqrt{3}}\Lambda_{2}
-\frac{1}{4\sqrt{3}}\left(3m_{1}+m_{2}-1+3m_{3}+m_{4}-m_{6}\right)\right]X^{-}_{m_{2}-1}\nonumber\\
&& \qquad\qquad
+ \frac{1}{2\sqrt{2}}m_{3}X^{-}_{m_{3}-1,m_{4}+1}
+ \frac{1}{\sqrt{6}}m_{4}X^{-}_{m_{4}-1,m_{6}+1}  \nonumber\\
&& \qquad\qquad -
\frac{1}{8\sqrt{3}}{m_{4}(m_{4}-1)}X^{-}_{m_{4}-2,m_{5}+1},\nonumber\\
&&\rho (E_{3}) X^{-}_{(m)} =
\frac{1}{16\sqrt{6}}{m_{1}m_{4}(m_{4}-1)}X^{-}_{m_{1}-1,m_{2}+1,m_{4}-2}\nonumber\\
&& \qquad\qquad -
\frac{1}{96\sqrt{3}}{m_{1}m_{4}(m_{4}-1)(m_{4}-2)}X^{-}_{m_{1}-1,m_{3}+1,m_{4}-3} \nonumber\\
&& \qquad\qquad -
\frac{1}{16\sqrt{3}}m_{1}m_{4}m_{6}\left(\Lambda_{2}-\frac{m_{6}-1}{4\sqrt{3}}\right)X^{-}_{m_{1}-1,m_{4}-1,m_{6}-1} \nonumber\\
&& \qquad\qquad -
\frac{1}{8\sqrt{2}}m_{1}m_{5}\left[\Lambda_{1}+{\sqrt{3}}\Lambda_{2}
- \frac{1}{2}\left(m_{4}+m_{5}-1+m_{6}\right)\right]X^{-}_{m_{1}-1,m_{5}-1}\nonumber\\
&& \qquad\qquad
-\frac{1}{48\sqrt{6}}{m_{2}(m_{2}-1)(m_{2}-2)}X^{-}_{m_{1}+1,m_{2}-3}\nonumber\\
&& \qquad\qquad +
\frac{1}{8\sqrt{2}}m_{2}m_{4}\left[\Lambda_{1}+\frac{1}{\sqrt{3}}\Lambda_{2}
-\frac{1}{6}\left(2m_{2}+m_{4}-3+3m_{5}+m_{6}\right)\right]X^{-}_{m_{2}-1,m_{4}-1}\nonumber\\
&& \qquad\qquad +
\frac{1}{8}m_{2}m_{5}X^{-}_{m_{2}-1,m_{5}-1,m_{6}+1} \nonumber\\
&& \qquad\qquad +
\frac{1}{192}{m_{2}(m_{2}-1)}{m_{4}(m_{4}-1)}X^{-}_{m_{2}-2,m_{3}+1,m_{4}-2}\nonumber\\
&& \qquad\qquad -
\frac{1}{16\sqrt{6}}{m_{2}(m_{2}-1)m_{5}}X^{-}_{m_{2}-2,m_{4}+1,m_{5}-1}\nonumber\\
&& \qquad\qquad +
\frac{1}{48}{m_{2}(m_{2}-1)}m_{6}\left(\Lambda_{2} \; -\frac{m_{6}-1}{4\sqrt{3}}\right)X^{-}_{m_{2}-2,m_{6}-1}\nonumber\\
&& \qquad\qquad +
\frac{1}{2}m_{3}\left[\Lambda_{1}-\frac{1}{4}\left(m_{1}+m_{2}+m_{3}-1+m_{4}+m_{5}\right)\right]X^{-}_{m_{3}-1}
\nonumber\\&& \qquad\qquad -
\frac{1}{8\sqrt{3}}{m_{4}(m_{4}-1)}X^{-}_{m_{4}-2,m_{6}+1}\nonumber\\
&& \qquad\qquad +
\frac{1}{48\sqrt{6}}{m_{4}(m_{4}-1)(m_{4}-2)}X^{-}_{m_{4}-3,m_{5}+1},\nonumber\\
&&\rho(E_{4}) X^{-}_{(m)} =
- \frac{1}{8\sqrt{3}}{m_{2}(m_{2}-1)}X^{-}_{m_{1}+1,m_{2}-2}\nonumber\\
&& \qquad\qquad +
\frac{1}{24\sqrt{2}}{m_{2}m_{4}(m_{4}-1)}X^{-}_{m_{2}-1,m_{3}+1,m_{4}-2}\nonumber\\
&& \qquad\qquad
- \frac{1}{4\sqrt{3}}m_{2}m_{5}X^{-}_{m_{2}-1,m_{4}+1,m_{5}-1}\nonumber\\
&& \qquad\qquad +
\frac{1}{6\sqrt{2}}m_{2}m_{6}\left(\Lambda_{2}- \frac{m_{6}-1}{4\sqrt{2}}\right)X^{-}_{m_{2}-1,m_{6}-1}\nonumber\\
&& \qquad\qquad -\frac{1}{2\sqrt{2}}m_{3}X^{-}_{m_{2}+1,m_{3}-1} + \frac{1}{2\sqrt{2}}m_{5}X^{-}_{m_{5}-1,m_{6}+1}\nonumber\\
&& \qquad\qquad +
\frac{1}{4}m_{4}\left[\Lambda_{1}+\frac{1}{\sqrt{3}}\Lambda_{2}-\frac{1}{6}\left(4m_{2}+m_{4}-1+3m_{5}+m_{6}\right)
\right]X^{-}_{m_{4}-1},\nonumber\\
&&\rho(E_{5}) X^{-}_{(m)} = \frac{1}{2\sqrt{2}}m_{3}X^{-}_{m_{1}+1,m_{3}-1}-\frac{1}{8\sqrt{3}}{m_{4}(m_{4}-1)}X^{-}_{m_{2}+1,m_{4}-2}\nonumber\\
&& \qquad\qquad +
\frac{1}{24\sqrt{6}}{m_{4}(m_{4}-1)(m_{4}-2)}X^{-}_{m_{3}+1,m_{4}-3}\nonumber\\
&& \qquad\qquad +
\frac{1}{4\sqrt{6}}m_{4}m_{6}\left(\Lambda_{2}-
\frac{m_{6}-1}{4\sqrt{3}}\right)X^{-}_{m_{4}-1,m_{6}-1}\nonumber\\
&&\qquad\qquad +
\frac{1}{4}m_{5}\left[\Lambda_{1}+{\sqrt{3}}\Lambda_{2}
- \frac{1}{2}({m_{4}+m_{5}-1+m_{6}})\right]X^{-}_{m_{5}-1}, \nonumber\\
&&\rho(E_{6}) X^{-}_{(m)} =
-\frac{1}{2\sqrt{2}}m_{2}X^{-}_{m_{1}+1,m_{2}-1}-\frac{1}{\sqrt{6}}m_{4}X^{-}_{m_{2}+1,m_{4}-1}-\frac{1}{2\sqrt{2}}m_{5}X^{-}_{m_{4}+1,m_{5}-1}\nonumber\\
&& \qquad\qquad +
\frac{1}{8\sqrt{3}}{m_{4}(m_{4}-1)}X^{-}_{m_{3}+1,m_{4}-2}
+\frac{1}{2\sqrt{3}}m_{6}\left(\Lambda_{2}-
\frac{m_{6}-1}{4\sqrt{3}}\right)X^{-}_{m_{6}-1},\nonumber\\
&&\rho(H_{1}) X^{-}_{(m)}
=\left[\Lambda_{1}-\frac{1}{4}(m_{1}+m_{2}+2m_{3}+m_{4}+m_{5})\right]
X^{-}_{(m)}\equiv M_1 X^{-}_{(m)} ,\nonumber\\
&&\rho (H_{2}) X^{-}_{(m)}
=\left[\Lambda_{2}+\frac{1}{4\sqrt{3}}(3m_{1}+m_{2}-m_{4}-3m_{5}-2m_{6})\right]
X^{-}_{(m)}\equiv M_2 X^{-}_{(m)}.
\end{eqnarray}

It is clear from the last two equations of Eq. (\ref{eq:dlambda}) that $X_{(m)}^{-}$ is the common eigenvector of $H_1$ and $H_2$, with their respective eigenvalues being $M_1$ and $M_2$, i.e., the weight of $X_{(m)}^{-}$ is $\mathbf{M}\equiv (M_1,M_2)$. The weight diagram of $d_{\mathbf{\Lambda}}$ is the same as Fig. \ref{fig:4diagrams}(c) with the highest weight being $\mathbf{\Lambda}$ instead of $\mathbf{0}$.

By making use of Eq. (\ref{eq:bosonrela}), we can immediately obtain from the matrix elements (\ref{eq:dlambda}) of $d_{\mathbf{\Lambda}}$ the inhomogeneous six-boson realization of $G_2$:
\begin{eqnarray}\label{6bosonRe}
&&B (E_{-1}) = a_{1}^{\dag}, \nonumber\\
&&B (E_{-2}) = a_{2}^{\dag} ,\nonumber\\
&&B (E_{-3}) = a_{3}^{\dag} ,\nonumber\\
&&B (E_{-4}) = a_{4}^{\dag} - \frac{1}{2\sqrt{2}} a_{3}^{\dag} a_{2} ,\nonumber\\
&&B (E_{-5}) = a_{5}^{\dag} + \frac{1}{2\sqrt{2}} a_{3}^{\dag} a_{1} ,\nonumber\\
&&B (E_{-6}) = a_{6}^{\dag} - \frac{1}{2\sqrt{2}} a_{2}^{\dag} a_{1}  - \frac{1}{\sqrt{6}} a_{4}^{\dag} a_{2}  +\frac{1}{8\sqrt{3}} a_{3}^{\dag} a_{2}^{2} -
\frac{1}{2\sqrt{2}} a_{5}^{\dag} a_{4}  ,\nonumber\\
&&B (E_{1}) = \frac{1}{4} \left[ \Lambda_{1}- \sqrt{3} \Lambda_{2} - \frac{1}{2}(\hat{n}_{1}+ \hat{n}_{2} + \hat{n}_{3} - \hat{n}_{5} - \hat{n}_{6}) \right] a_{1}\nonumber\\
&& \qquad\qquad +\frac{1}{2\sqrt{2}} a_{6}^{\dag} a_{2} - \frac{1}{8\sqrt{3}} a_{4}^{\dag} a_{2}^{2} + \frac{1}{48\sqrt{6}} a_{3}^{\dag} a_{2}^{3}- \frac{1}{8}a_{5}^{\dag} a_{2} a_{4} - \frac{1}{2\sqrt{2}} a_{5}^{\dag} a_{3} ,\nonumber\\
&&B (E_{2})  = - \frac{1}{4\sqrt{3}} a_{2}^{\dag} a_{1} a_{4} + \frac{1}{16\sqrt{6}}a_{3}^{\dag} a_{1} a_{4}^{2} -
\frac{1}{8} a_{4}^{\dag} a_{1}a_{5}+ \frac{1}{4\sqrt{6}} \left( \Lambda_{2}-\frac{1}{4\sqrt{3}}\hat{n}_{6} \right) a_{1} a_{6}
\nonumber\\ && \qquad\qquad  +
\frac{1}{4} \left[ \Lambda_{1} -\frac{1}{\sqrt{3}}\Lambda_{2} - \frac{1}{4 \sqrt{3}} (3\hat{n}_{1} + \hat{n}_{2} +3 \hat{n}_{3} + \hat{n}_{4} - \hat{n}_{6}) \right] a_{2} \nonumber\\
&& \qquad\qquad +\frac{1}{2\sqrt{2}} a_{4}^{\dag} a_{3} +\frac{1}{\sqrt{6}}a_{6}^{\dag} a_{4} - \frac{1}{8\sqrt{3}}a_{5}^{\dag} a_{4}^{2},\nonumber\\
&&B (E_{3})  =\frac{1}{16\sqrt{6}}a_{2}^{\dag}a_{1}a_{4}^{2} - \frac{1}{96\sqrt{3}}a_{3}^{\dag}a_{1}a_{4}^3 - \frac{1}{16\sqrt{3}}\left(\Lambda_{2}-\frac{1}{4\sqrt{3}}\hat{n}_{6}\right)a_{1}a_{4}a_{6} \nonumber\\
&& \qquad\qquad -
\frac{1}{8\sqrt{2}}\left[\Lambda_{1}+{\sqrt{3}}\Lambda_{2}-\frac{1}{2}(\hat{n}_{4}+\hat{n}_{5}+\hat{n}_{6})
\right]a_{1} a_{5}\nonumber\\
&& \qquad\qquad -\frac{1}{48\sqrt{6}} a_{1}^{\dag} a_{2}^{3}+
\frac{1}{8\sqrt{2}}\left[\Lambda_{1}+\frac{1}{\sqrt{3}}\Lambda_{2}
-\frac{1}{6}\left(2\hat{n}_{2}+\hat{n}_{4}+3\hat{n}_{5}+\hat{n}_{6}\right)\right]a_{2}a_{4}\nonumber\\
&& \qquad\qquad +
\frac{1}{8}a_{6}^{\dag} a_{2} a_{5} +\frac{1}{192}a_{3}^{\dag}  a_{2}^{2}a_{4}^{2}
-\frac{1}{16\sqrt{6}}a_{4}^{\dag} a_{2}^{2} a_{5}
+\frac{1}{48}\left(\Lambda_{2}
-\frac{\hat{n}_{6}}{4\sqrt{3}}\right) a_{2}^{2} a_{6}\nonumber\\
&& \qquad\qquad +
\frac{1}{2} \left[ \Lambda_{1}-\frac{1}{4}(\hat{n}_{1}+\hat{n}_{2}+\hat{n}_{3}+\hat{n}_{4}+\hat{n}_{5}) \right] a_{3} - \frac{1}{8\sqrt{3}} a_{6}^{\dag} a_{4}^{2} + \frac{1}{48\sqrt{6}} a_{5}^{\dag} a_{4}^{3},\nonumber\\
&&B(E_{4}) = -\frac{1}{8\sqrt{3}}a_{1}^{\dag}a_{2}^{2} + \frac{1}{24\sqrt{2}}a_{3}^{\dag} a_{2}a_{4}^{2} \;
-\frac{1}{4\sqrt{3}}a_{4}^{\dag}a_{2}a_{5} \nonumber\\
&& \qquad\qquad +
\frac{1}{6\sqrt{2}}\left(\Lambda_{2}-
\frac{1}{4\sqrt{3}}\hat{n}_{6}\right)a_{2}a_{6}
-\frac{1}{2\sqrt{2}}a_{2}^{\dag}a_{3} +
\frac{1}{2\sqrt{2}}a_{6}^{\dag}a_{5} \nonumber\\
&& \qquad\qquad +
 \frac{1}{4} \left[ \Lambda_{1}+\frac{1}{\sqrt{3}}\Lambda_{2} - \frac{1}{6}(4\hat{n}_{2}+\hat{n}_{4}+3\hat{n}_{5}+\hat{n}_{6})\right]a_{4},\nonumber\\
&&B(E_{5})  = \frac{1}{2\sqrt{2}}a_{1}^{\dag}a_{3}
-\frac{1}{8\sqrt{3}}a_{2}^{\dag}a_{4}^{2}
+\frac{1}{24\sqrt{6}}a_{3}^{\dag}a_{4}^{3} \nonumber\\
&& \qquad\qquad + \frac{1}{4\sqrt{6}} \left( \Lambda_{2}- \frac{1}{4\sqrt{3}}\hat{n}_{6} \right) a_{4} a_{6} +\left[\frac{1}{4}\Lambda_{1}+\frac{\sqrt{3}}{4}\Lambda_{2}-
\frac{1}{8}\left(\hat{n}_{4}+\hat{n}_{5}+\hat{n}_{6}\right)\right]a_{5},\nonumber\\
&&B(E_{6}) =
-\frac{1}{2\sqrt{2}}a_{1}^{\dag}a_{2}-\frac{1}{\sqrt{6}}a_{2}^{\dag}a_{4}-\frac{1}{2\sqrt{2}}a_{4}^{\dag}a_{5}
+ \frac{1}{8\sqrt{3}}a_{3}^{\dag}a_{4}^{2}
+ \frac{1}{2\sqrt{3}} \left( \Lambda_{2}- \frac{1}{4 \sqrt{3}} \hat{n}_{6} \right) a_{6},\nonumber\\
&&B(H_{1})=\Lambda_{1}-\frac{1}{4}\left(\hat{n}_{1}+\hat{n}_{2}+2\hat{n}_{3}+\hat{n}_{4}+\hat{n}_{5}\right),\nonumber\\
&&B (H_{2})=\Lambda_{2}+\frac{1}{4\sqrt{3}}\left(3\hat{n}_{1}+\hat{n}_{2}-\hat{n}_{4}-3\hat{n}_{5}-2\hat{n}_{6}\right).
\end{eqnarray}

The boson realization of $H_1$ and $H_2$ is dependent on the particle number operators $\hat{n}_i$ $(i=1,2,\dots,6)$ only. Notice that $\hat{n}_6$ and $\hat{n}_3$ disappear in $B(H_{1})$ and $B(H_{2})$, respectively.

\section{TWO SERIES OF REPRESENTATIONS ON QUOTIENT SPACES OF $\Omega_{-}$ INDUCED FROM THE ELEMENTARY REPRESENTATION}\label{sec:extremal}

In this section the extremal vectors defining invariant subspaces will be calculated, and all corresponding elementary representations will be determined. On quotient spaces with respect to these invariant subspaces, two series of representations induced from $d_{\mathbf{\Lambda}}$ will be discussed in detail, which are infinite-dimensional indecomposable, infinite-dimensional irreducible or finite-dimensional irreducible. The calculation of new IHBRs will be carried out in the same manner as in the last section.

\subsection{Extremal vectors and the series of elementary representations}\label{extremals}

According to the definition given in Ref. [44], a vector $Y\in \Omega_{-}$ is called an extremal vector of the elementary representation $d_{\mathbf{\Lambda}}$, if it satisfies
\begin{equation}{\label{eq:extremal}}
\left\{\begin{array}{ll}
\rho(H_{j})Y = M_{j}Y,\quad M_{j} \in \mathbb{C},\;j=1,2,\\
\rho(E_i)Y = 0, \quad i=1,2,\dots,6.
\end{array}\right.
\end{equation}
Therefore an extremal vector $Y$ (with its weight being $\mathbf{M}$) is the highest weight vector of the subspace $$I_Y\equiv\Omega_{-}Y: \{E_{-1}^{m_1}E_{-2}^{m_2}\dots E_{-6}^{m_6}Y|m_i \in \mathbb{Z}^{+},i=1,2,\ldots,6\},$$ which is obviously an invariant subspace (ideal) of $\Omega_{-}$. The representation on $I_Y$, which may be induced from $d_{\mathbf{\Lambda}}$, is also an elementary representation of $G_2$, here denoted by $d_{\mathbf{M}}$. It is clear from Eq. (\ref{eq:extremal}) that the identity vector $\mathbf{1}$ is an extremal vector of $d_{\mathbf{\Lambda}}$.

Bernshtein, Gel'fand, and Gel'fand\cite{BGG} gave the necessary and sufficient condition that the module $d_{\mathbf{\Lambda}}$ contains $d_{\mathbf{M}}$ (known as BGG theorem): there exist a sequence of positive roots $\gamma_1,\gamma_2,\dots,\gamma_k$ $(k\in \mathbb{N})$, which satisfy
\begin{eqnarray}\label{eq:MofY}
\left\{\begin{array}{ll}
    \mathbf{M}+\mathbf{R}=S_{\gamma_{k}}\dots S_{\gamma_2}S_{\gamma_1}(\mathbf{\Lambda} + \mathbf{R}),\\
    2(S_{\gamma_{i-1}}\dots S_{\gamma_2}S_{\gamma_1}(\mathbf{\Lambda} + \mathbf{R}),\gamma_i)/(\gamma_i,\gamma_i)\in \mathbb{N},\quad i=1,2,\ldots,k,
    \end{array}
\right.
\end{eqnarray}
where $S_{\gamma_i}$ is the Weyl reflection for $\gamma_i$, $S_{\gamma_0}\equiv 1$, and $\mathbf{R}$ is the half sum of all positive roots. Eq. (\ref{eq:MofY}) gives the weights of all possible extremal vectors of $d_{\mathbf{\Lambda}}$. Gruber {\it et al.}\cite{Gruber7} put forward an algorithm to construct all extremal vectors. Suppose that first one knows an extremal vector $Z_{1}\in \Omega_{-}$, which defines an ideal $\Omega_{-}Z_{1}\subset\Omega_{-}$. Let $\mathbf{M}_{1}$ denote the weight of $Z_{1}$, then $\mathbf{M}_{1}$ is the highest weight of a new elementary representation $d_{\mathbf{M}_{1}}$ defined on $\Omega_{-}Z_{1}$. If $d_{\mathbf{M}_{1}}$ contains another extremal vector $Z_{2}\in \Omega_{-}Z_{1}$ with the weight $\mathbf{M}_{2}$, one can construct another elementary representation $d_{\mathbf{M}_{2}}$ on the ideal $\Omega_{-}Z_{2}Z_{1} \subset \Omega_{-}Z_{1}$, with $\mathbf{M}_{2}$ being its highest weight. $\Omega_{-}Z_{2}Z_{1}$ is also an ideal of $\Omega_{-}$, and therefore the vector $Z_{2}Z_{1}$ is the extremal vector defining $\Omega_{-}Z_{2}Z_{1}$ with respect to $\Omega_{-}$. Repeating this procedure, one may find the $i$th extremal vector $Y_{i}\equiv Z_{i}Z_{i-1}\dots Z_{1}$, and the corresponding ideal $I_{Y_i}\equiv \Omega_{-}Z_{i}Z_{i-1}\dots Z_{1}$. Thus, all ideals form an ideal chain $I_{Y_i}\subset I_{Y_{i-1}}\subset\dots\subset I_{Y_1}\subset\Omega_{-}$ naturally.

Now consider an elementary representation $d_\mathbf{\Lambda}$ of $G_2$, with $\mathbf{\Lambda}$ being dominant, i.e., $\langle\mathbf{\Lambda}, \alpha_i\rangle \equiv 2(\mathbf{\Lambda},\alpha_i)/(\alpha_i,\alpha_i)$ are nonnegative integers for both simple roots $\alpha_1$ and $\alpha_6$. From the explicit matrix elements of $d_{\mathbf{\Lambda}}$ given by Eq. (\ref{eq:dlambda}), one can find that for the simple root vectors $E_{-1}^{m_1}$ and $E_{-6}^{m_6}$, most terms in Eq. (\ref{eq:dlambda}) vanish, and the second condition of Eq. (\ref{eq:extremal}) becomes
\begin{eqnarray}{\label{eq:y1}}
\left\{\begin{array}{ll}
\rho(E_1)E_{-1}^{m_{1}} &= m_1\left(\frac{\Lambda_{1}}{4}-\frac{\sqrt{3}\Lambda_{2}}{4}-\frac{m_{1}-1}{8}\right)E_{-1}^{m_{1}-1} \\
&= m_1\frac{(\alpha_{1},\alpha_{1})}{2}\left[\frac{2(\mathbf{\Lambda},\alpha_{1})}{(\alpha_{1},\alpha_{1})}-(m_1-1)\right]
E_{-1}^{m_{1}-1}=0,\\
\rho(E_i)E_{-1}^{m_1} &= 0, \quad i=2,3,\ldots,6 \end{array}\right.
\end{eqnarray}
for $E_{-1}^{m_1}$, and
\begin{eqnarray}{\label{eq:y2}}
\left\{\begin{array}{ll}\rho(E_i)E_{-6}^{m_6} &=0, \quad i=1,2,\ldots,5, \\
\rho(E_6)E_{-6}^{m_6} &=m_6(\frac{\Lambda_{2}}{2\sqrt{3}}-\frac{m_{6}-1}{24})E_{-6}^{m_6-1}\\
&= m_6\frac{(\alpha_{6},\alpha_{6})}{2}\left[\frac{2(\mathbf{\Lambda},\alpha_{6})}{(\alpha_{6},\alpha_{6})}-(m_6-1)\right]E_{-6}^{m_6-1}=0
\end{array}\right.
\end{eqnarray}
for $E_{-6}^{m_6}$.
Solving (\ref{eq:y1}) and (\ref{eq:y2}) gives
\begin{eqnarray}{\label{eq:domin}}
&\left\{\begin{array}{ll}
m_1=\frac{2(\mathbf{\Lambda},\alpha_{1})}{(\alpha_{1},\alpha_{1})}+1=p+1=\frac{2(\mathbf{\Lambda}+\mathbf{R},\alpha_{1})}{(\alpha_{1},\alpha_{1})}\equiv P, \\
m_6=\frac{2(\mathbf{\Lambda},\alpha_{6})}{(\alpha_{6},\alpha_{6})}+1=q+1=\frac{2(\mathbf{\Lambda}+\mathbf{R},\alpha_{6})}{(\alpha_{6},\alpha_{6})}\equiv Q. \end{array}\right.
\end{eqnarray}
Since $\mathbf{\Lambda}$ is dominant, $p,q$ are nonnegative integers and $P,Q$ are positive integers.

\begin{figure}[!hbp]
\begin{center}
\includegraphics[width=0.7\textwidth]{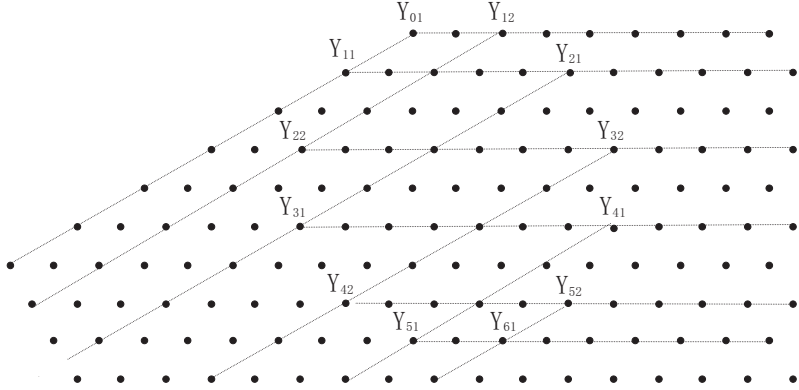}
\caption{The twelve extremal vectors $Y_{ik}$ on the weight diagram of the elementary representation $d_{\mathbf{\Lambda}}$. $Y_{01}\equiv \mathbf{1}$ is the highest weight vector of  $d_{\mathbf{\Lambda}}$. Extremal vectors below $\mathbf{1}$ can be categorized into different layers: $Y_{ik}$ ($i=1,2,\dots,6$) belong to the $i$th layer below $\mathbf{1}$. Extremal vectors in the same layer are not higher nor lower than each other, while $Y_{ik}$ is always higher than $Y_{jl}$ if $i<j$.}{\label{fig:extrem}}
\end{center}
\end{figure}

Let $Y_{01}$ denote the extremal vector $\mathbf{1}$. Here we use the double-subscript because there are two series of extremal vectors, and therefore two ideal chains (as we can see below). The first subscript is used to label the layers, and the second one the ideal chains. It follows from Eqs. (\ref{eq:y1}) - (\ref{eq:domin}) that we obtain in $\Omega_{-}$ two extremal vectors $Y_{11}=E_{-1}^{P}$ and $Y_{12}=E_{-6}^{Q}$, with their respective weights being
\begin{eqnarray}
  \mathbf{M}_{{11}}&=& \mathbf{\Lambda} - P\alpha_{1}=\mathbf{\Lambda} - \frac{2(\mathbf{\Lambda}+\mathbf{R},\alpha_{1})}{(\alpha_{1},\alpha_{1})}\alpha_{1}= S_{\alpha_1}(\mathbf{\Lambda} +\mathbf{R}) - \mathbf{R},  \\
  \mathbf{M}_{{12}}&=&  \mathbf{\Lambda} - Q\alpha_{6}=\mathbf{\Lambda} - \frac{2(\mathbf{\Lambda}+\mathbf{R},\alpha_{6})}{(\alpha_{6},\alpha_{6})}\alpha_{6}= S_{\alpha_6}(\mathbf{\Lambda} +\mathbf{R}) - \mathbf{R}. {\label{eq:weightofY}}
\end{eqnarray}
$Y_{11}$ and $Y_{12}$ are the first layer of extremal vectors below $\mathbf{1}$ (see Fig. \ref{fig:extrem}).


Now we use $Y_{11}$ and $\mathbf{M}_{{11}}$ in substitution for $Y_{01}$ and ${\mathbf{\Lambda}}$ respectively, then Eqs. (\ref{eq:y1}) - (\ref{eq:weightofY}) give the extremal vector $Z_{21}=E_{-6}^{Q'}$ in $\Omega_{-}{Y_{11}}$, and  Eq. (\ref{eq:domin}) gives $P'=3P+Q$. The corresponding extremal vector with respect to $\Omega_{-}$ is therefore $Y_{21}=E_{-6}^{Q'}Y_{11}=E_{-6}^{3P+Q}E_{-1}^{P}$. Similar arguments for $Y_{12}$ give another extremal vector $Y_{22}=E_{-1}^{P'}Y_{12}=E_{-1}^{P+Q}E_{-6}^{Q}$. Obviously, $Y_{21}$ and $Y_{22}$ are products of two simple root vectors, and they are the second layer of extremal vectors below $\mathbf{1}$.

Repeating this procedure, we can obtain all twelve extremal vectors of $G_2$ as follows:
\begin{eqnarray}{\label{eq:allextre}}
&&\qquad\qquad\qquad\qquad Y_{01} \equiv  \mathbf{1}, \nonumber\\
&&  Y_{11} = E_{-1}^{P}Y_{01},
\qquad\qquad\qquad\qquad\;\;
    Y_{12} = E_{-6}^{Q}Y_{01},\nonumber\\
&&  Y_{21} = E_{-6}^{3P+Q}Y_{11},
\qquad\qquad\qquad\quad\;
    Y_{22} = E_{-1}^{P+Q}Y_{12},\nonumber\\
&&  Y_{31} = E_{-1}^{2P+Q}Y_{21},
\qquad\qquad\qquad\quad\;
    Y_{32} = E_{-6}^{3P+2Q}Y_{22}, \nonumber\\
&&  Y_{41} = E_{-6}^{3P+2Q}Y_{31},\qquad\qquad\qquad\;\;\;
    Y_{42} = E_{-1}^{2P+Q}Y_{32},\nonumber\\
&&  Y_{51} = E_{-1}^{P+Q}Y_{41},\qquad\qquad\qquad\quad\;\;
    Y_{52} = E_{-6}^{3P+Q}Y_{42},\nonumber\\
&& \qquad\qquad\quad Y_{61}  = E_{-6}^{Q}Y_{51}=E_{-1}^{P}Y_{52}.
\end{eqnarray}

Although here we start from the highest extremal vector $\mathbf{1}$, it is easy to verify that starting from any extremal vector, we may recover the same twelve extremal vectors. Fig. \ref{fig:extrem} shows the positions of these twelve extremal vectors on the weight diagram of $d_{\mathbf{\Lambda}}$. $Y_{ik}$ $(i=1,2,\dots,6, k=1,2)$ belong to the $i$th layer below $\mathbf{1}$, which are products of $i$ simple root vectors. Especially, for the $6$th layer there is only one extremal vector $Y_{61}$, because its two forms in Eq. (\ref{eq:allextre}) are in fact equal (This kind of equivalence has been generally proved by Verma\cite{Verma}). Each of these extremal vectors defines an ideal $I_{ik}\equiv \Omega_{-}Y_{ik}$ of $\Omega_{-}$, which carries an elementary representation $d_{\mathbf{M}_{ik}}$.

If some extremal vector $Y_{ik}$ belongs to the higher layer than $Y_{jl}$ (i.e. $i < j$), then its weight $\mathbf{M}_{ik}$ is higher than the weight $ \mathbf{M}_{jl}$ of $Y_{jl}$, and their corresponding ideals satisfy $I_{jl}\subset I_{ik}$, which can be shown by BGG theorem, 
therefore, we have $I_{ik}+I_{jl}=I_{ik}$. In order to find new ideals with respect to an elementary representation, we need turn to the sums of subspaces defined by the extremal vectors belonging to the same layer
\begin{equation}\label{newideal}
   I_{j1}+I_{j2}, \quad j= 1,2,\ldots,5.
\end{equation}
These five subspaces, which have no highest weight vectors, are new ideals of spaces $I_{ik}$ $(i<j)$.

Using these extremal vectors and ideals, one can investigate various new representations on the quotient spaces of $I_{ik}$, i.e., on $I_{ik}/I_{jl}$ $(i<j)$ and $I_{ik}/(I_{j1}+I_{j2})$ $(i<j)$. If their matrix elements can be explicitly determined, then one may obtain new oscillator realizations of $G_2$.

\subsection{Representations on $I_{ik}/I_{jl}$ and a new five-boson realization}
Since all spaces $I_{jl}$ with $j>i$ are ideals of $I_{ik}$, it is easy to determine that there are 61 different quotient spaces $I_{ik}/I_{jl}$ in all. We denote by $d_{I_{ik}/I_{jl}}$ the representation on the quotient space $I_{ik}/I_{jl}$. Except that the extremal vectors $Y_{51}$ and $Y_{52}$ have only one extremal vector ($Y_{61}$) below them, the other extremal vectors $Y_{ik}$ ($i\leq 4,k=1,2$) have more than two different extremal vectors below them. Therefore, $d_{I_{51}/I_{61}}$ and $d_{I_{52}/I_{61}}$ are infinite-dimensional irreducible, while the other 59 representations $d_{I_{ik}/I_{jl}}$ $(i\leq 4, i<j)$ are infinite-dimensional indecomposable.

Matrix elements of the representation $d_{I_{ik}/I_{jl}}$ $(i<j)$ can be obtained by imposing ${\Omega_{-}Y_{jl}}=0$ in ${d_{\mathbf{M}_{ik}}}$. Note that the extremal vectors given by Eq. (\ref{eq:allextre}) are not in the standard order (see Eq. (\ref{eq:standardbasisofmaster})). To get the basis of $I_{ik}/I_{jl}$ explicitly, one must reexpress the equations ${\Omega_{-}Y_{jl}}=0$ into the standard ordered form by making use of Eqs. (\ref{eq:commurela1})-(\ref{eq:commurela2}). For the general case, this re-ordering is too complicated to be done and unenlightening. However, for the special case $d_{I_{01}/I_{12}}$, the equations ${\Omega_{-}Y_{12}}=0$ are already in the standard order, therefore the explicit matrix elements can be simply obtained from $d_{\mathbf{\Lambda}}$ by setting $E_{-1}^{m_{1}} E_{-2}^{m_{2}} E_{-3}^{m_{3}}E_{-4}^{m_{4}} E_{-5}^{m_{5}} E_{-6}^{Q+m_{6}}=0$ for all $m_{i}\in \mathbb{Z}^{+}$. Thus, the representation space of $d_{I_{01}/I_{12}}$ is
\begin{equation}\label{quotientbasis}
V_{I_{01}/I_{12}}:\{E_{-1}^{m_{1}} E_{-2}^{m_{2}} E_{-3}^{m_{3}}E_{-4}^{m_{4}} E_{-5}^{m_{5}} E_{-6}^{m_{6}}| m_{6}<Q, m_{i}\in \mathbb{Z}^{+}\}.
\end{equation}
If $q=2(\Lambda,\alpha_6)/(\alpha_6,\alpha_6)=0$, i.e., $\Lambda_{2}=0$, then $Q=1$, $V_{I_{01}/I_{12}}$ becomes
\begin{equation}\label{5tensorbasis}
V_{I_{01}/I_{12}}':\{E_{-1}^{m_{1}} E_{-2}^{m_{2}} E_{-3}^{m_{3}}E_{-4}^{m_{4}} E_{-5}^{m_{5}} | m_{i}\in \mathbb{Z}^{+}\}.
\end{equation}
By discarding all terms including  $m_{6}$ in $d_{\mathbf{\Lambda}}$ given by Eq. ({\ref{eq:dlambda}}), one may get the explicit matrix elements of the representation $d_{I_{01}/I_{12}}$, which are not given here. Furthermore, with the help of Eq. (\ref{eq:bosonrela}), one may obtain an inhomogeneous five-boson realization of $G_2$:
\begin{eqnarray*}
B (E_{-1}) & = & a_{1}^{\dag},\\
B (E_{-2}) & = & a_{2}^{\dag}, \\
B (E_{-3}) & = & a_{3}^{\dag}, \\
B (E_{-4}) & = & a_{4}^{\dag} - \frac{1}{2\sqrt{2}} a_{3}^{\dag} a_{2}, \\
B (E_{-5}) & = & a_{5}^{\dag} + \frac{1}{2\sqrt{2}} a_{3}^{\dag} a_{1}, \\
B (E_{-6}) & = &  - \frac{1}{2\sqrt{2}} a_{2}^{\dag} a_{1}  - \frac{1}{\sqrt{6}} a_{4}^{\dag} a_{2}
  - \frac{1}{8\sqrt{3}} a_{3}^{\dag} a_{2}^{2} - \frac{1}{2\sqrt{2}} a_{5}^{\dag} a_{4},\\
B (E_{1}) & = & \frac{1}{48\sqrt{6}} a_{3}^{\dag} a_{2}^{3} - \frac{1}{8\sqrt{3}} a_{4}^{\dag} a_{2}^{2}
   - \frac{1}{8}a_{5}^{\dag} a_{2} a_{4} - \frac{1}{2\sqrt{2}} a_{5}^{\dag} a_{3} \\
&& + \frac{1}{4} \left[ \Lambda_{1} - \frac{1}{2}\left(\hat{n}_{1}+ \hat{n}_{2} + \hat{n}_{3} - \hat{n}_{5} \right)\right]a_{1},\\
B (E_{2})  & = &
 \frac{1}{2\sqrt{2}} a_{4}^{\dag} a_{3} + \frac{1}{16\sqrt{6}}a_{3}^{\dag} a_{1} a_{4}^{2} - \frac{1}{4\sqrt{3}} a_{2}^{\dag} a_{1} a_{4} - \frac{1}{8} a_{4}^{\dag} a_{1}a_{5} \\
&& + \frac{1}{4} \left[ \Lambda_{1} - \frac{1}{6}\left(3\hat{n}_{1} +2\hat{n}_{2} +3 \hat{n}_{3} + \hat{n}_{4}\right) \right] a_{2} - \frac{1}{8\sqrt{3}}a_{5}^{\dag} a_{4}^{2}, \\
B (E_{3})  & = & - \frac{1}{48\sqrt{6}} a_{1}^{\dag} a_{2}^{3} + \frac{1}{8\sqrt{2}}\left[\Lambda_{1} -\frac{1}{6}\left(2\hat{n}_{2}+\hat{n}_{4}
 +3 \hat{n}_{5}\right)\right]a_{2}a_{4}  - \frac{1}{16\sqrt{6}}a_{4}^{\dag} a_{2}^{2} a_{5} \\
&& + \frac{1}{192}a_{3}^{\dag} a_{2}^{2} a_{4}^{2} + \frac{1}{2} \left[ \Lambda_{1} - \frac{1}{4}\left(\hat{n}_{1}+\hat{n}_{2}+\hat{n}_{3}
 +\hat{n}_{4}+\hat{n}_{5}\right)\right] a_{3} \\
&& + \frac{1}{48\sqrt{6}} a_{5}^{\dag} a_{4}^{3} + \frac{1}{16\sqrt{6}}a_{2}^{\dag}a_{1}a_{4}^{2} - \frac{1}{96\sqrt{3}}a_{3}^{\dag}a_{1}a_{4}^3 \\
&&  - \frac{1}{8\sqrt{2}}\left[\Lambda_{1}-\frac{1}{2}(\hat{n}_{4}+\hat{n}_{5}) \right]a_{1}a_{5} ,\\
B (E_{4}) & = & -\frac{1}{8\sqrt{3}}a_{1}^{\dag}a_{2}^{2} + \frac{1}{24\sqrt{2}}a_{3}^{\dag} a_{2} a_{4}^{2} - \frac{1}{4\sqrt{3}}a_{4}^{\dag}a_{2}a_{5} \\
&& - \frac{1}{2\sqrt{2}}a_{2}^{\dag}a_{3} + \frac{1}{4} \left[ \Lambda_{1}-\frac{1}{6}\left(4\hat{n}_{2}+\hat{n}_{4}+3\hat{n}_{5}\right)\right]a_{4},\\
B (E_{5})  & = & \frac{1}{2\sqrt{2}}a_{1}^{\dag}a_{3} - \frac{1}{8\sqrt{3}}a_{2}^{\dag}a_{4}^{2} + \frac{1}{24\sqrt{6}}a_{3}^{\dag}a_{4}^{3}
 +\frac{1}{4}\left[\Lambda_{1}- \frac{1}{2}(\hat{n}_{4}+\hat{n}_{5})\right]a_{5},\\
B (E_{6}) & = & -\frac{1}{2\sqrt{2}}a_{1}^{\dag}a_{2} -\frac{1}{\sqrt{6}}a_{2}^{\dag}a_{4}-\frac{1}{2\sqrt{2}}a_{4}^{\dag}a_{5} +\frac{1}{8\sqrt{3}}a_{3}^{\dag}a_{4}^{2},\\
B (H_{1}) & = & \Lambda_{1}-\frac{1}{4}\left(\hat{n}_{1} +\hat{n}_{2}+2\hat{n}_{3}+\hat{n}_{4}+\hat{n}_{5}\right), \\
B (H_{2}) & = & \frac{1}{4\sqrt{3}}\left(3\hat{n}_{1}+\hat{n}_{2} -\hat{n}_{4}-3\hat{n}_{5}\right).
\end{eqnarray*}

If one chooses another basis of $\Omega_{-}$ in a different order with $E_{-1}$ on the right (for example, $E_{-6}^{m_6}E_{-5}^{m_5}\dots E_{-1}^{m_1}$) instead of the standard order in Eq. (\ref{factorspace}), then using the same method one can get another indecomposable representation $d_{I_{01}/I_{11}}$ involving five generators $E_{-i}$ ($i=2,\dots,6$), and correspondingly, one can obtain a new inhomogeneous five-boson realization.

\subsection{Representations on the quotient spaces ${I_{ik}}/({I_{j1}}+{I_{j2}})$ and a three-fermion realization of fundamental representation $(0,1)$}
We denote by $d_{I_{ik}/({I_{j1}}+{I_{j2}})}$ the representation on the quotient space ${I_{ik}}/({I_{j1}}+{I_{j2}})$. Owing to the fact that the spaces $I_{j1}+I_{j2}$ ($j=1,2,\dots,5$) are ideals of the spaces $I_{ik}$ with $i<j$, there are in all 25 different quotient spaces ${I_{ik}}/({I_{j1}}+{I_{j2}})$. For the nine cases of the nearest-neighbor layers the representations $d_{I_{j-1,k}/({I_{j1}}+{I_{j2}})}$ are irreducible, because $I_{j1}+I_{j2}$ are the largest ideal of $I_{j-1,k}$. For the remaining cases the corresponding representations $d_{I_{ik}/({I_{j1}}+{I_{j2}})}$ ($i<j-1$) are indecomposable. These representations are infinite-dimensional except for $d_{I_{01}/({I_{11}}+{I_{12}})}$.

The matrix elements of $d_{I_{ik}/({I_{j1}}+{I_{j2}})}$ may be determined from $d_{\mathbf{M}_{ik}}$ by setting in $I_{ik}$:
\begin{eqnarray}\label{ElmentsOfQuotientII}
\left\{\begin{array}{ll}
\Omega_{-}Y_{j1} =0,\\
\Omega_{-}Y_{j2} =0.
\end{array} \right.
\end{eqnarray}
For the infinite-dimensional representations $d_{I_{ik}/({I_{j1}}+{I_{j2}})}$ ($i\neq 0$), the PBW basis of the subspaces $\Omega_{-}Y_{jl}$ ($l=1,2$) here are not in the standard order, however, the re-ordering is too complicated to be done. Whereas, for $d_{I_{01}/({I_{11}}+{I_{12}})}$ the re-ordering becomes possible since it is finite-dimensional. The representation $d_{I_{01}/({I_{11}}+{I_{12}})}$ can also be labeled by Dynkin symbol $(p,q)$ (where $p=P-1$, $q=Q-1$), which is related to $\mathbf{\Lambda}$ by Eq. (\ref{eq:domin}). As an example, we now calculate the simplest case of the fundamental representation $(0,1)$, which is seven-dimensional.

\begin{figure}[!hbp]
\begin{center}
\includegraphics[width=0.4\textwidth]{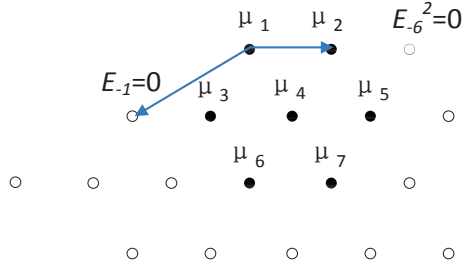}
\caption{Dots are the seven weights ($\mu_{i}$, $i=1,\dots,7$) of the weight diagram of the fundamental representation (0,1) of $G_2$, and $\textrm{dim } V(\mu_{i})=1$. Circles denote the weights of the other weight spaces of $d_{\mathbf{\Lambda}}$ (here $\mathbf{\Lambda}=(\frac{1}{4},\frac{1}{4\sqrt{3}})$), which are void in the quotient space $I_{01}/(I_{11} + I_{12})$. $I_{01}$ is the whole representation space of $d_{\mathbf{\Lambda}}$, $
I_{11}=\Omega_{-}E_{-1}$ and $I_{12}=\Omega_{-}E_{-6}^2$.}{\label{fig:p0q1}}
\end{center}
\end{figure}

For (0,1), it follows from Eq. (\ref{eq:domin}) that $\mathbf{\Lambda}$ is $(\frac{1}{4},\frac{1}{4\sqrt{3}})$ and $P=1,Q=2$, i.e., $Y_{11}=E_{-1}^{1},Y_{12}=E_{-6}^{2}$. The weight diagram of $(0,1)$ is shown in Fig. \ref{fig:p0q1}. We denote by $\mu_i$ ($i=1,2,\dots,7$) the seven weights of $(0,1)$, and by $V(\mu_{i})$ and $\Omega_{-}(\mu_{i})$ the corresponding weight spaces of $(0,1)$ and $d_{Y_{01}}$ respectively. The standard ordered PBW bases of $\Omega_{-}(\mu_{i})$ read respectively
\begin{eqnarray}\label{basisOfOmega01}
\Omega_{-}(\mu_{1}): & & \{\mathbf{1}\} ,\nonumber\\
\Omega_{-}(\mu_{2}): & & \{E_{-6}\}     ,\nonumber\\
\Omega_{-}(\mu_{3}): & & \{E_{-2},E_{-1}E_{-6}\} ,\nonumber\\
\Omega_{-}(\mu_{4}): & & \{E_{-4},E_{-2}E_{-6},E_{-1}E_{-6}^2\} ,\nonumber\\
\Omega_{-}(\mu_{5}): & & \{E_{-5},E_{-4}E_{-6},E_{-2}E_{-6}^2,E_{-1}E_{-6}^3\}  ,\nonumber \\
\Omega_{-}(\mu_{6}): & & \{E_{-3},E_{-2}E_{-4},E_{-1}E_{-5},E_{-2}^2E_{-6},E_{-1}E_{-4}E_{-6},
                       E_{-1}E_{-2}E_{-6}^2,E_{-1}^2E_{-6}^3\}   ,\nonumber\\
\Omega_{-}(\mu_{7}): & & \left\{E_{-4}^2,E_{-2}E_{-5},E_{-3}E_{-6},E_{-2}E_{-4}E_{-6},E_{-1}E_{-5}E_{-6},\right.\nonumber\\
                       & & \left.E_{-2}^2E_{-6}^2,E_{-1}E_{-4}E_{-6}^2,E_{-1}E_{-2}E_{-6}^3,E_{-1}^2E_{-6}^3\right\}.
\end{eqnarray}

On weight spaces $\Omega_{-}(\mu_{1})$ and $\Omega_{-}(\mu_{2})$, the equations
\begin{eqnarray}\label{imposingfor01}
   \left\{\begin{array}{ll}\Omega_{-}Y_{12}=\{E_{-1}^{m_{1}} E_{-2}^{m_{2}} E_{-3}^{m_{3}}E_{-4}^{m_{4}} E_{-5}^{m_{5}} E_{-6}^{m_{6}}\}E_{-6}^{2}=0 ,\\
   \Omega_{-}Y_{11}=\{E_{-1}^{m_{1}} E_{-2}^{m_{2}} E_{-3}^{m_{3}}E_{-4}^{m_{4}} E_{-5}^{m_{5}} E_{-6}^{m_{6}}\}E_{-1}=0 \end{array}\right.
\end{eqnarray}
give no constraint relations, so that we get $V(\mu_1)=\Omega_{-}(\mu_1)$ and $V(\mu_2)=\Omega_{-}(\mu_2)$. On $\Omega_{-}(\mu_{3})$, the condition $\Omega_{-}Y_{12}=0$ gives no constraints, while $\Omega_{-}Y_{11}=0$ requires $E_{-6}Y_{11}=E_{-6}E_{-1}=0$, i.e., $E_{-1}E_{-6}-1/(2\sqrt{2})E_{-2}=0$, which shows that the two basis elements $E_{-1}E_{-6}$ and $E_{-2}$ are linearly dependent, therefore, for $(0,1)$ the corresponding weight space is $V(\mu_3)=\{E_{-2}=2\sqrt{2}E_{-1}E_{-6}\}$, which is one-dimensional. It follows from Eq. (\ref{imposingfor01}) that by the same method we can determine the constraint relations among basis elements for $\Omega_{-}(\mu_4),\Omega_{-}(\mu_5),\Omega_{-}(\mu_6)$, and $\Omega_{-}(\mu_7)$. Thus the bases of the seven weight spaces $V(\mu_i)$ of $(0,1)$ are respectively




\begin{eqnarray}\label{basisOfV01}
  V(\mu_{1}):& & \{\mathbf{1}\} ,\nonumber\\
  V(\mu_{2}):& & \{E_{-6}\}     ,\nonumber\\
  V(\mu_{3}):& & \{E_{-2}=2\sqrt{2}E_{-1}E_{-6}\} ,\nonumber\\
  V(\mu_{4}):& & \{E_{-4}=2\sqrt{6}E_{-2}E_{-6}\} , \nonumber\\
  V(\mu_{5}):& & \{E_{-4}E_{-6}=\frac{1}{6\sqrt{2}}E_{-5}\}  ,\nonumber \\
  V(\mu_{6}):& & \{E_{-2}E_{-4}=3\sqrt{2}E_{-3}=-\frac{2}{3}E_{-1}E_{-5}=2\sqrt{6}E_{-2}^2E_{-6}=-4\sqrt{2}E_{-1}E_{-4}E_{-6}\},  \nonumber\\
  V(\mu_{7}):& & \{E_{-2}E_{-4}E_{-6}=-\frac{1}{4\sqrt{6}}E_{-4}^2=\frac{1}{6\sqrt{2}}E_{-2}E_{-5}=\frac{1}{6\sqrt{2}}E_{-3}E_{-6}                       \}. \nonumber\\
\end{eqnarray}
All of them are one-dimensional.

Eq. (\ref{basisOfV01}) shows that for (0,1) if we choose the basis
\begin{equation}\label{basisOfV}
V_{(0,1)}:\{\mathbf{1},E_{-6},E_{-2},E_{-4},E_{-4}E_{-6},E_{-2}E_{-4},E_{-2}E_{-4}E_{-6}\},
\end{equation}
then we get the representation space with its PBW basis consisting of only three independent generators with powers no larger than one. This makes it possible to construct a three-fermion realization with respect to (0,1).

The basis elements of $V_{(0,1)}$ given by Eq. (\ref{basisOfV}) may be written in the convenient form
\begin{equation*}
    \{X_{(m_2,m_4,m_6)}=E_{2}^{m_2}E_{4}^{m_4}E_{6}^{m_6}|m_2,m_4,m_6=0,1\}.
\end{equation*}
The nonzero actions of $E_{\pm i}$ ($i=1,\dots,6$) upon the basis elements in Eq. (\ref{basisOfV}) can be obtained by Eqs. (\ref{commurela1}) - (\ref{commurela2}). Using the method very parallel to the bosonic case (see Eqs. (\ref{eq:Fbas}) - (\ref{eq:gamma})), the fermion Fock space
$$\mathcal{F}_{f}: \{|m_2,m_4,m_6\rangle= (f_2^{\dag})^{m_2}(f_4^{\dag})^{m_4}(f_6^{\dag})^{m_6}|0\rangle\}$$
may be defined, where $f_{i}^{\dag}$ and $f_{i}$ ($i=2,4,6$) are the creation and annihilation operators of the $i$th fermion, and $|0\rangle$ denotes the vacuum state of $\mathcal{F}_{f}$. The anti-commutation relations are
\begin{eqnarray}\label{fermioncomm}
\{f_{i},f_{i}^{\dag}\} = \delta_{ij},\quad
\{f_{i},f_{j}\} = \{f_{i}^{\dag}, f_{j}^{\dag}\}=0.
\end{eqnarray}
The following relations may be used to simplify calculations
\begin{equation}\label{fermionconstaints}
\left\{\begin{array}{ll}
\hat{n}^{f}_{i}=f_{i}^{\dag}f_i=(f_{i}^{\dag}f_{i})^2=(\hat{n}^{f}_{i})^2,\\
(1-\hat{n}^{f}_{i})^2=1-\hat{n}^{f}_{i},\\
f_{i} f_{i}^{\dag} f_{i} = (1-f_{i}^{\dag} f_{i})f_{i} = f_{i},\\
f_{i}^{\dag} f_{i} f_{i}^{\dag} = f_{i}^{\dag}(1-f_{i}^{\dag} f_{i})=f_{i}^{\dag},
\end{array}\right.
\end{equation}
where $\hat{n}_{i}^{f}$ is the particle number operator of the $i$th fermion.

The actions of $E_{-1}$ on the basis of $V_{(0,1)}$ are
\begin{equation}\label{actionofEm1}
   \left\{\begin{array}{ll} \phi({E_{-1}})X_{(0,0,1)}=\frac{1}{2\sqrt{2}}X_{(1,0,0)},\\
   \phi({E_{-1}})X_{(0,1,1)}=-\frac{1}{4\sqrt{2}}X_{(1,1,0)},\\
   \phi(E_{-1})X_{(m_2,m_4,m_6)}=0 \quad\textrm{for other five basis elements}.
\end{array}\right.
\end{equation}
By making use of the $\delta$-function, Eq. (\ref{actionofEm1}) can be written in a more compact form
\begin{eqnarray}\label{rho1}
 \phi(E_{-1})X_{(m_2,m_4,m_6)}=     (\frac{1}{2\sqrt{2}}\delta_{0,m_2}\delta_{0,m_4}\delta_{1,m_6}
-\frac{1}{4\sqrt{2}}\delta_{0,m_2}\delta_{1,m_4}\delta_{1,m_6})X_{(m_2+1,m_4,m_6-1)}.
\end{eqnarray}
Since $m_i$ can take 0 or 1 only, all terms involving $\delta_{0,m_i}X_{(m_i-1)}$ and $\delta_{1,m_i}X_{(m_i+1)}$ vanish. For the non-zero terms there exist the following correspondence relations:
\begin{eqnarray}\label{3fermionrela}
\left\{\begin{array}{ll}
  \delta_{0,m_i}X_{(m_i+1)}\longmapsto f_{i}^{\dag}(1-f_{i}^{\dag}f_{i})|{m_i}\rangle=f_{i}^{\dag}|{m_i}\rangle,\\
  \delta_{0,m_i}X_{(m_i)}\longmapsto (1-f_{i}^{\dag}f_{i})|{m_i}\rangle=(1-\hat{n}^{f}_{i})|{m_i}\rangle,\\
  \delta_{1,m_i}X_{(m_i)}\longmapsto f_{i}^{\dag}f_{i}|{m_i}\rangle=\hat{n}^{f}_{i}|{m_i}\rangle,\\
  \delta_{1,m_i}X_{(m_i-1)}\longmapsto f_{i}(f_{i}^{\dag}f_{i})|{m_i}\rangle=f_{i}|{m_i}\rangle.
\end{array}\right.
\end{eqnarray}
Then from Eq. (\ref{rho1}) one gets the fermion realization of $E_{-1}$
\begin{eqnarray}\label{fermion1}
 F(E_{-1}) = \frac{1}{2\sqrt{2}}f_{2}^{\dag}(1-f_{4}^{\dag}f_{4})f_6
 +\frac{1}{4\sqrt{2}}f_{2}^{\dag}f_{4}^{\dag}f_{4}f_6=\frac{1}{2\sqrt{2}}(1-\frac{1}{2}\hat{n}_{4}^{f})f_{2}^{\dag}f_6.
\end{eqnarray}
Realizations of the remaining generators of $G_2$ can be calculated by the same method. They are
\begin{eqnarray}\label{3fermionrealization}
 F(E_{-2})&=&(1+\hat{n}^{f}_{4}\hat{n}^{f}_{6}-\hat{n}^{f}_{6})f_{2}^{\dag}+\frac{1}{2\sqrt{6}}(1-\hat{n}^{f}_{2})f_{4}^{\dag}f_{6}\nonumber,\\
 F(E_{-3})&=&3\sqrt{2}(1+\hat{n}^{f}_{6})f_{2}^{\dag}f_{4}^{\dag},\nonumber\\
 F(E_{-4})&=& (1-\frac{1}{2}\hat{n}^{f}_{2}-\frac{1}{2}\hat{n}^{f}_{2}\hat{n}^{f}_{6}) f_{4}^{\dag}+4\sqrt{6}\hat{n}^{f}_{4}f_{2}^{\dag}f_{6}^{\dag},\nonumber\\
 F(E_{-5})&=& 6\sqrt{2}f_{4}^{\dag}f_{6}^{\dag},\nonumber\\
 F(E_{-6})&=& (1-\hat{n}^{f}_{2}+\hat{n}^{f}_{4}+\hat{n}^{f}_{2}\hat{n}^{f}_{4})f_{6}^{\dag}
 +\frac{1}{2\sqrt{6}}(1-\hat{n}^{f}_{6})f_{2}f_{4}^{\dag},\nonumber\\
 F(E_1)&=& -\frac{1}{2\sqrt{2}}(1+\hat{n}^{f}_{4})f_{2}f_{6}^{\dag},\nonumber\\
 F(E_2)&=& \frac{1}{24} (1+\hat{n}^{f}_{4}-\hat{n}^{f}_{6})f_{2}-\frac{1}{\sqrt{6}}(1-\hat{n}^{f}_{2})f_{4}f_{6}^{\dag},\nonumber\\
 F(E_3)&=& -\frac{1}{24\sqrt{2}}(1-\frac{1}{2}\hat{n}^{f}_{6})f_{2}f_{4},\nonumber\\
 F(E_4)&=& \frac{1}{12}(1-\frac{1}{2}\hat{n}^{f}_{6}-\frac{1}{2}\hat{n}^{f}_{2}\hat{n}^{f}_{6})f_{4}-\frac{1}{48\sqrt{6}}\hat{n}^{f}_{4}f_{2}f_{6},\nonumber\\
 F(E_5)&=& -\frac{1}{48\sqrt{2}}f_{4}f_{6},\nonumber\\
 F(E_6)&=&\frac{1}{24}(1-\hat{n}^{f}_{2}+\frac{1}{2}\hat{n}^{f}_{2}\hat{n}^{f}_{4})f_{6}-\frac{1}{\sqrt{6}}(1-\hat{n}^{f}_{6})f_{2}^{\dag}f_{4},\nonumber\\
 F(H_{1})&=& \frac{1}{4}(1-\hat{n}^{f}_{2}-\hat{n}^{f}_{4}),\nonumber\\
 F(H_{2})&=& \frac{1}{4\sqrt{3}}(1+\hat{n}^{f}_{2}-\hat{n}^{f}_{4}-2\hat{n}^{f}_{6}).
\end{eqnarray}
It is easy to verify with the help of Eqs. (\ref{fermioncomm}) and (\ref{fermionconstaints}) that they satisfy the commutation relations of $G_2$. Hence from the fundamental representation $(0,1)$, we obtain a new three-fermion realization, which is different from the fermion realizations based upon two Lie algebra chains $so(8)\supset G_2$ and $G_2 \supset su(3)$.\cite{Dictionary}

\section{SUMMARY AND DISCUSSIONS}
In this paper we have obtained the explicit matrix elements of the master representation $\rho$ of $G_2$, which is defined on the space $\Omega$ of the universal enveloping algebra of $G_2$, and discussed various representations induced from $\rho$, such as the representations on the invariant subspaces of $\Omega$ and the representations on the quotient spaces with respect to these invariant subspaces. Particularly, we have investigated in detail the elementary representations of $G_2$, which are defined on the subspace $\Omega_{-}$ (generated by $E_{-i}$ only) of $\Omega$ and have an highest weight, and from which we have realized $G_2$ by six boson pairs. For the elementary representation with a fixed dominate $\mathbf{\Lambda}$ being its highest weight, we have obtained all twelve extremal vectors in explicit form, each of them defines an ideal $I_{ik}$ of $\Omega_{-}$, which carries an elementary sub-representation. With the help of these extremal vectors we have investigated the representations on the quotient spaces of $\Omega_{-}$ with fewer generators, and consequently constructed the five-boson realization from $d_{I_{01}/I_{12}}$ and the three-fermion realization from the special case of $d_{I_{01}/(I_{11}+I_{12})}=(0,1)$.

If one makes use of the following corresponding relations between the boson operators and the differential operators
$$
 a_{i}^{\dag}\Leftrightarrow x_i,\quad a_{i}\Leftrightarrow \frac{\partial}{\partial x_i},
$$
where $x_i$ and $\frac{\partial}{\partial x_i}$ satisfy the commutation relations
\begin{equation*}
    \left[\frac{\partial}{\partial x_i},x_j\right]=\delta_{ij},\quad \left[\frac{\partial}{\partial x_i},\frac{\partial}{\partial x_j}\right]=\left[x_i,x_j\right]=0,
\end{equation*}
then one can obtain immediately the corresponding inhomogeneous differential realizations of $G_2$ from the six-boson or five-boson realizations obtained in this paper.

If one chooses as the PBW basis for $\Omega$  the set of elements in a different order, for example
\begin{equation*}
\Omega :\{ E_{1}^{n_{1}} E_{2}^{n_{2}} E_{3}^{n_{3}} E_{4}^{n_{4}} E_{5}^{n_{5}} E_{6}^{n_{6}} E_{-1}^{m_{1}} E_{-2}^{m_{2}} E_{-3}^{m_{3}}
E_{-4}^{m_{4}} E_{-5}^{m_{5}} E_{-6}^{m_{6}}H_{1}^{k_1}H_{2}^{k_2}\},
\end{equation*}
then one can study in the same manner the elementary representations on $\Omega_{+}:$ $\{ E_{1}^{n_{1}} E_{2}^{n_{2}} E_{3}^{n_{3}} E_{4}^{n_{4}} E_{5}^{n_{5}} E_{6}^{n_{6}} \}$, the extremal vectors, and the representations on various ideals or quotient spaces with respect to $\Omega_{+}$. These representations have their respective lowest weights, whereas the representations on the spaces with respect to $\Omega_{-}$ discussed in this paper have the highest weights. Similarly, one can obtain the corresponding boson or fermion realizations.

Similar procedure can also be used to investigate the exceptional Lie superalgebra $G(3)$, which is underway.

\section*{ACKNOWLEDGEMENTS}
This work is supported by the State Key Basic Research Development Programs (Grant No. 2009CB929402).

\newpage

\appendix
\section{The matrix elements of the master representation $\rho$ of $G_2$}
Here we do not give the concrete values of the coefficients $N_{\alpha,\beta}$ of $G_2$, thus, if we choose the other basis (for example, the irreducible tensor basis), we can easily get the new master representation by simply replacing the symbols of the basis and the coefficients in $\rho$. Using the convenient abbreviations: $X_{(mnk)} \equiv X({\ldots,m_{i},\ldots})$ and $X_{m_i+k} \equiv X(\ldots,m_{i}+k,\ldots)$ in $\Omega$, the explicit matrix elements of $\rho$ are
\setlength\arraycolsep{1pt}
\begin{eqnarray*}
&&\rho (E_{-1})X_{(mnk)} =X_{m_{1}+1}, \\
&&\rho (E_{-2})X_{(mnk)} =X_{m_{2}+1} ,\\
&&\rho (E_{-3})X_{(mnk)} =X_{m_{3}+1}, \\
&&\rho (E_{-4})X_{(mnk)} =X_{m_{4}+1}  +  m_{2} N_{-4, -2} X_{m_{2}-1,m_{3}+1}, \\
&&\rho (E_{-5})X_{(mnk)} =X_{m_{5}+1} +  m_{1} N_{-5,-1} X_{m_{1}-1,m_{3}+1} ,\\
&&\rho (E_{-6})X_{(mnk)} =X_{m_{6}+1} + m_{1} N_{-6, -1} X_{m_{1}-1 m_{2}+1} +  m_{2} N_{-6, -2} X_{m_{2}-1 m_{4}+1} \\
&& \qquad\qquad + \frac{1}{2}m_{2}(m_{2}-1) N_{-6, -2} N_{-4,-2}X_{m_{2}-1,m_{3}+1} +
m_{4} N_{-6, -4} X_{m_{4}-1 m_{5}+1},\\
&&\rho (E_{1})X_{(mnk)} = X_{n_{1}+1}\\
&& \qquad\qquad +
m_{1}\left[\frac{1}{2}(\alpha_{1},-\alpha_{1}){(m_{1}-1)}+ \sum_{i=2}^{6}(\alpha_{1},-\alpha_{i})m_{i} + \sum_{i=1}^{6}(\alpha_{1},\alpha_{i})n_{i}
 \right]X_{m_{1}-1}\\
&& \qquad\qquad +
m_{1}\alpha_{1}^{(1)}X_{m_{1}-1,k_{1}+1 }
+ m_{1} \alpha_{1}^{(2)}X_{m_{1}-1,k_{2}+1}\\
&& \qquad\qquad +  m_{2}N_{1,-2}X_{m_{2}-1,m_{6}+1}
+ \frac{1}{2}{m_{2}(m_{2}-1)} N_{1,-2} N_{-6,-2} X_{m_{2}-2,m_{4}+1}\\
&& \qquad\qquad +
\frac{1}{6}{m_{2}(m_{2}-1)(m_{2}-2)}N_{1,-2}
N_{-6,-2} N_{-4,-2} X_{m_{2}-2,m_{3}+1} \\
&& \qquad\qquad +
m_{2} m_{4} N_{1,-2}
N_{-6,-4}X_{m_{2}-1,m_{4}-1,m_{5}+1}\\
&& \qquad\qquad +
m_{3} N_{1,-3} X_{m_{3}-1,m_{5}+1 } ,\\
&&\rho (E_{2}) X_{(mnk)} = X_{n_{2}+1}\\
&& \qquad\qquad  +
m_{2}\left[ m_{1} N_{2,-1} N_{6,-2}+\frac{1}{2}(\alpha_{2},-\alpha_{2}){(m_{2}-1)}+
\sum_{i=3}^{6}(\alpha_{2},-\alpha_{i})
m_{i} \right. \\ && \qquad\qquad\qquad\qquad
+ \left. \sum_{i=1}^{6}(\alpha_{2},\alpha_{i}) n_{i} \right]X_{m_{2}-1}\\
&& \qquad\qquad+
m_{1}N_{2,-1}X_{m_{1}-1,n_{6}+1} \\
&& \qquad\qquad + m_{1}m_{4}N_{2,-1}N_{6,-4}X_{m_{1}-1,m_{2}+1,m_{4}-1} \\
&& \qquad\qquad +
\frac{1}{2}{m_{1}m_{4}(m_{4}-1)}N_{2,-1}N_{6,-4}N_{-4,-2}X_{m_{1}-1,m_{3}+1,m_{4}-2} \\
&& \qquad\qquad +
m_{1}m_{5}N_{2,-1}N_{6,-5}X_{m_{1}-1,m_{4}+1,m_{5}-1} \\
&& \qquad\qquad + m_{1}m_{6}N_{2,-1}
\left[\frac{1}{2}(\alpha_{6},-\alpha_{6}){(m_{6}-1)}+ \sum_{i=1}^{6}(\alpha_{6},\alpha_{i})n_{i} \right] X_{m_{1}-1,m_{6}-1} \\
&& \qquad\qquad +
m_{1}m_{6}N_{2,-1}\alpha_{6}^{(1)}X_{m_{1}-1,m_{6}-1,k_{1}+1}\\
&& \qquad\qquad +
m_{1}m_{6}N_{2,-1}\alpha_{6}^{(2)}X_{m_{1}-1,m_{6}-1,k_{2}+1} \\
&& \qquad\qquad +
m_{1}n_{1}N_{2,-1}N_{6,1}X_{m_{1}-1,n_{1}-1,n_{2}+1} \\
&& \qquad\qquad +
m_{1}n_{2}N_{2,-1}N_{6,2}X_{m_{1}-1,n_{2}-1,n_{4}+1} \\
&& \qquad\qquad +
\frac{1}{2}{m_{1}n_{2}(n_{2}-1)}N_{2,-1}N_{6,2}N_{4,2}X_{m_{1}-1,n_{2}-2,n_{3}+1} \\
&& \qquad\qquad +
m_{1}n_{4}N_{2,-1}N_{6,4}X_{m_{1}-1,n_{4}-1,n_{5}+1} \\
&& \qquad\qquad+  m_{2}\alpha_{2}^{(1)}
X_{m_{2}-1,k_{1}+1} +  m_{2}\alpha_{2}^{(2)}
X_{m_{2}-1,k_{2}+1}\\
&& \qquad\qquad
+  m_{3}N_{2,-3}X_{m_{3}-1,m_{4}+1}\\
&& \qquad\qquad + m_{4}N_{2,-4}X_{m_{4}-1,m_{6}+1} \\
&& \qquad\qquad +
\frac{1}{2}{m_{4}(m_{4}-1)}N_{2,-4}N_{-6,-4}X_{m_{4}-2,m_{5}+1}\\
&& \qquad\qquad + m_{6}N_{2,-6}X_{m_{6}-1,n_{1}+1}, \\
&&\rho (E_{3}) X_{(mnk)} = X_{n_{3}+1}\\
&& \qquad\qquad +  m_{1}N_{3,-1}X_{m_{1}-1,n_{5}+1}\\
&& \qquad\qquad +
m_{1}m_{3}N_{3,-1}N_{5,-3}X_{m_{3}-1}\\
&& \qquad\qquad +
m_{1}m_{4}N_{3,-1}N_{5,-4}X_{m_{1}-1,m_{4}-1,n_{6}+1}\\
&& \qquad\qquad +
\frac{1}{2}{m_{1}m_{4}(m_{4}-1)}N_{3,-1}N_{5,-4}N_{6,-4}X_{m_{1}-1,m_{2}+1,m_{4}-2}\\
&& \qquad\qquad +
\frac{1}{3}{m_{1}m_{4}(m_{4}-1)(m_{4}-2)}N_{3,-1}N_{5,-4}N_{6,-4}N_{-4,-2}X_{m_{1}-1,m_{3}+1,m_{4}-3}\\
&& \qquad\qquad +
m_{1}m_{4}m_{5}N_{3,-1}N_{5,-4}N_{6,-5}X_{m_{1}-1,m_{5}-1} \\
&& \qquad\qquad +
m_{1}m_{4}m_{6}N_{3,-1}N_{5,-4}\left[\frac{1}{2}(\alpha_{6},-\alpha_{6}){(m_{6}-1)}+\sum_{i=1}^{6}(\alpha_{6},\alpha_{i})n_{i}
\right]X_{m_{1}-1,m_{4}-1,m_{6}-1}\\
&& \qquad\qquad +
m_{1}m_{4}m_{6}N_{3,-1}N_{5,-4}\alpha_{6}^{(1)}X_{m_{1}-1,m_{4}-1,m_{6}-1,k_{1}+1}\\
&& \qquad\qquad +
m_{1}m_{4}m_{6}N_{3,-1}N_{5,-4}\alpha_{6}^{(2)}X_{m_{1}-1,m_{4}-1,m_{6}-1,k_{2}+1}\\
&& \qquad\qquad +
m_{1}m_{4}n_{1}N_{3,-1}N_{5,-4}N_{6,1}X_{m_{1}-1,m_{4}-1,n_{1}-1,n_{2}+1}\\
&& \qquad\qquad +
m_{1}m_{4}n_{2}N_{3,-1}N_{5,-4}N_{6,2}X_{m_{1}-1,m_{4}-1,n_{2}-1,n_{4}+1}\\
&& \qquad\qquad +
\frac{1}{2}{m_{1}m_{4}n_{2}(n_{2}-1)}N_{3,-1}N_{5,-4}N_{6,2}N_{4,2}X_{m_{1}-1,m_{4}-1,n_{2}-2,n_{3}+1} \\
&& \qquad\qquad +
m_{1}m_{4}n_{4}N_{3,-1}N_{5,-4}N_{6,4}X_{m_{1}-1,m_{4}-1,n_{4}-1,n_{5}+1}\\
&& \qquad\qquad +
m_{1}m_{5}N_{3,-1}\left[\frac{1}{2}(\alpha_{5},-\alpha_{5}){(m_{5}-1)}+(\alpha_{5},-\alpha_{6})m_{6}+\sum_{i=1}^{6}(\alpha_{5},\alpha_{i})n_{i}
\right]X_{m_{1}-1,m_{5}-1}\\
&& \qquad\qquad +
m_{1}m_{5}N_{3,-1}\alpha_{5}^{(1)}X_{m_{1}-1,m_{5}-1,k_{1}+1}+ m_{1}m_{5}N_{3,-1}\alpha_{5}^{(2)}X_{m_{1}-1,m_{5}-1,k_{2}+1}\\
&& \qquad\qquad +
m_{1}m_{6}N_{3,-1}N_{5,-6}X_{m_{1}-1,m_{6}-1,n_{4}+1} \\
&& \qquad\qquad +
\frac{1}{2}{m_{1}m_{6}(m_{6}-1)}N_{3,-1}N_{5,-6}N_{4,-6}X_{m_{1}-1,m_{6}-2,n_{2}+1}\\
&& \qquad\qquad +
\frac{1}{6}{m_{1}m_{6}(m_{6}-1)(m_{6}-2)}N_{3,-1}N_{5,-6}N_{4,-6}N_{2,-6}X_{m_{1}-1,m_{6}-3,n_{1}+1}\\
&& \qquad\qquad +
m_{1}m_{6}n_{2}N_{3,-1}N_{5,-6}N_{4,2}X_{m_{1}-1,m_{6}-1,n_{2}-1,n_{3}+1}\\
&& \qquad\qquad +  m_{1}n_{1}N_{3,-1}N_{5,1}X_{m_{1}-1,n_{1}-1,n_{3}+1}
\\&& \qquad\qquad +  m_{2}N_{3,-2}X_{m_{2}-1,n_{4}+1}\\
&& \qquad\qquad +
\frac{1}{2}{m_{2}(m_{2}-1)}N_{3,-2}N_{4,-2}X_{m_{2}-2,n_{6}+1}\\
&& \qquad\qquad +
\frac{1}{6}{m_{2}(m_{2}-1)(m_{2}-2)}N_{3,-2}N_{4,-2}N_{6,-2}X_{m_{1}+1,m_{2}-3}\\
&& \qquad\qquad +
m_{2}m_{3}N_{3,-2}N_{4,-3}X_{m_{3}-1} \\
&& \qquad\qquad +
m_{2}m_{4}N_{3,-2}\left[\frac{1}{2}(\alpha_{4},-\alpha_{4}){(m_{4}-1)}+\sum_{i=5}^{6}(\alpha_{4},-\alpha_{i})m_{i}
\right.\\
&& \qquad\qquad\qquad\qquad\qquad\qquad\qquad\qquad\qquad
\left.+ \sum_{i=1}^{6}(\alpha_{4},\alpha_{i})n_{i}\right]X_{m_{2}-1,m_{4}-1}\\
&& \qquad\qquad +
m_{2}m_{4}N_{3,-2}\alpha_{4}^{(1)}X_{m_{2}-1,m_{4}-1,k_{1}+1}+
m_{2}m_{4}N_{3,-2}\alpha_{4}^{(2)}X_{m_{2}-1,m_{4}-1,k_{2}+1}\\
&& \qquad\qquad +
m_{2}m_{5}N_{3,-2}N_{4,-5}X_{m_{2}-1,m_{5}-1,m_{6}+1} \\
&& \qquad\qquad +
m_{2}m_{6}N_{3,-2}N_{4,-6}X_{m_{2}-1,m_{6}-1,n_{2}+1}\\
&& \qquad\qquad +
\frac{1}{2}m_{2}{m_{6}(m_{6}-1)}N_{3,-2}N_{4,-6}N_{2,-6}X_{m_{2}-1,m_{6}-2,n_{1}+1}\\
&& \qquad\qquad +
m_{2}n_{2}N_{3,-2}N_{4,2}X_{m_{2}-1,n_{2}-1,n_{3}+1}\\
&& \qquad\qquad +
\frac{1}{2}{m_{2}(m_{2}-1)}m_{4}N_{3,-2}N_{4,-2}N_{6,-4}X_{m_{2}-1,m_{4}-1}\\
&& \qquad\qquad +
\frac{1}{4}{m_{2}(m_{2}-1)}{m_{4}(m_{4}-1)}N_{3,-2}N_{4,-2}N_{6,-4}N_{-4,-2}X_{m_{2}-2,m_{3}+1,m_{4}-2}\\
&& \qquad\qquad +
\frac{1}{2}{m_{2}(m_{2}-1)}m_{5}N_{3,-2}N_{4,-2}N_{6,-5}X_{m_{2}-2,m_{4}+1,m_{5}-1}\\
&& \qquad\qquad +
\frac{1}{2}{m_{2}(m_{2}-1)m_{6}}\left[\frac{1}{2}(\alpha_{6},-\alpha_{6}){(m_{6}-1)}+\sum_{i=1}^{6}(\alpha_{6},\alpha_{i})n_{i}
\right]X_{m_{2}-2,m_{6}-1}\\
&& \qquad\qquad +  \frac{1}{2}{m_{2}(m_{2}-1)m_{6}}N_{3,-2}N_{4,-2}
\alpha_{6}^{(1)}X_{m_{2}-2,m_{6}-1,k_{1}+1}\\
&& \qquad\qquad +
\frac{1}{2}{m_{2}(m_{2}-1)}m_{6}N_{3,-2}N_{4,-2}
\alpha_{6}^{(2)}X_{m_{2}-2,m_{6}-1,k_{2}+1}\\
&& \qquad\qquad +
\frac{1}{2}{m_{2}(m_{2}-1)}n_{1}N_{3,-2}N_{4,-2}N_{6,1}X_{m_{2}-2,n_{1}-1,n_{2}+1}\\
&& \qquad\qquad +
\frac{1}{2}{m_{2}(m_{2}-1)}n_{2}N_{3,-2}N_{4,-2}N_{6,2}X_{m_{2}-2,n_{2}-1,n_{4}+1}\\
&& \qquad\qquad +
\frac{1}{4}{m_{2}(m_{2}-1)}{n_{2}(n_{2}-1)}N_{3,-2}N_{4,-2}N_{6,2}N_{4,2}X_{m_{2}-2,n_{2}-2,n_{3}+1}\\
&& \qquad\qquad +
\frac{1}{2}{m_{2}(m_{2}-1)}n_{4}N_{3,-2}N_{4,-2}N_{6,4}X_{m_{2}-2,n_{4}-1,n_{5}+1}\\
&& \qquad\qquad +
m_{3}\left[\frac{1}{2}(\alpha_{3},-\alpha_{3}){(m_{3}-1)}+\sum_{i=4}^{6}(\alpha_{3},-\alpha_{i})m_{i}+ \sum_{i=1}^{6}(\alpha_{3},\alpha_{i})n_{i} \right] X_{m_{3}-1}\\
&& \qquad\qquad +
m_{3}\alpha_{3}^{(1)}X_{m_{3}-1,k_{1}+1}+ m_{3}\alpha_{3}^{(2)}X_{m_{3}-1,k_{2}+1} \\
&& \qquad\qquad + m_{4}N_{3,-4}X_{m_{4}-1,n_{2}+1}\\
&& \qquad\qquad +
\frac{1}{2}{m_{4}(m_{4}-1)}N_{3,-4}N_{2,-4}X_{m_{4}-2,m_{6}+1}\\
&& \qquad\qquad +
\frac{1}{6}{m_{4}(m_{4}-1)(m_{4}-2)}N_{3,-4}N_{2,-4}N_{-6,-4}X_{m_{4}-2,m_{5}+1} \\
&& \qquad\qquad +
m_{4}m_{6}N_{3,-4}N_{2,-6}X_{m_{4}-1,m_{6}-1,n_{1}+1}\\
&& \qquad\qquad +
m_{5}N_{3,-5}X_{m_{5}-1,n_{1}+1},\\
&&\rho(E_{4}) X_{(mnk)} =X_{n_{4}+1}\\
&& \qquad\qquad  +
m_{2}N_{4,-2}X_{m_{2}-1,n_{6}+1} \\
&& \qquad\qquad  +
\frac{1}{2}{m_{2}(m_{2}-1)}N_{4,-2}N_{6,-2}X_{m_{1}+1,m_{2}-2}\\
&& \qquad\qquad  +
m_{2}m_{4}N_{4,-2}N_{6,-4}X_{m_{4}-1}\\
&& \qquad\qquad  +
\frac{1}{2}{m_{2}m_{4}(m_{4}-1)}N_{4,-2}N_{6,-4}N_{-4,-2}X_{m_{2}-1,m_{3}+1,m_{4}-2}\\
&& \qquad\qquad +
m_{2}m_{5}N_{4,-2}N_{6,-5}X_{m_{2}-1,m_{4}+1,m_{5}-1} \\
&& \qquad\qquad +
m_{2}m_{6}N_{4,-2}\left[\frac{1}{2}(\alpha_{6},-\alpha_{6}){(m_{6}-1)}+\sum_{i=1}^{6}(\alpha_{6},\alpha_{i})n_{i}
\right]X_{m_{2}-1,m_{6}-1}\\
&& \qquad\qquad +
m_{2}m_{6}N_{4,-2}\alpha_{6}^{(1)}X_{m_{2}-1,m_{6}-1,k_{1}+1}+m_{2}m_{6}N_{4,-2}\alpha_{6}^{(2)}X_{m_{2}-1,m_{6}-1,k_{2}+1}\\
&& \qquad\qquad +
m_{2}n_{1}N_{4,-2}N_{6,1}X_{m_{2}-1,n_{1}-1,n_{2}+1}\\
&& \qquad\qquad +
m_{2}n_{2}N_{4,-2}N_{6,2}X_{m_{2}-1,n_{2}-1,n_{4}+1} \\
&& \qquad\qquad +
\frac{1}{2}{m_{2}n_{2}(n_{2}-1)}N_{4,-2}N_{6,2}N_{4,2}X_{m_{2}-1,n_{2}-2,n_{3}+1}\\
&& \qquad\qquad +
m_{2}n_{4}N_{4,-2}N_{6,4}X_{m_{2}-1,n_{4}-1,n_{5}+1} \\
&& \qquad\qquad +
m_{3}N_{4,-3}X_{m_{2}+1,m_{3}-1} \\
&& \qquad\qquad +
m_{4}\left[\frac{1}{2}(\alpha_{4},-\alpha_{4}){(m_{4}-1)}+\sum_{i=5}^{6}(\alpha_{4},-\alpha_{i})m_{i}+
\sum_{i=1}^{6}(\alpha_{4},\alpha_{i})n_{i} \right]X_{m_{4}-1}\\
&& \qquad\qquad +
m_{4}\alpha_{4}^{(1)}X_{m_{4}-1,k_{1}+1}\; + m_{4}\alpha_{4}^{(2)}X_{m_{4}-1,k_{2}+1}\\
&& \qquad\qquad + m_{5}N_{4,-5}X_{m_{5}-1,m_{6}+1}\\
&& \qquad\qquad + m_{6}N_{4,-6}X_{m_{6}-1,n_{2}+1}\\
&& \qquad\qquad + \frac{1}{2}{m_{6}(m_{6}-1)}N_{4,-6}N_{2,-6}X_{m_{6}-2,n_{1}+1} \\
&& \qquad\qquad + n_{2}N_{4,2}X_{n_{2}-1,n_{3}+1},\\
&&\rho(E_{5}) X_{(mnk)} =X_{n_{5}+1}\\
&& \qquad\qquad + m_{3}N_{5,-3}X_{m_{1}+1,m_{3}-1}\\
&& \qquad\qquad + m_{4}N_{5,-4}X_{m_{4}-1,n_{6}+1}\\
&& \qquad\qquad + \frac{1}{2}{m_{4}(m_{4}-1)}N_{5,-4}N_{6,-4}X_{m_{2}+1,m_{4}-2}\\
&& \qquad\qquad +
\frac{1}{3}{m_{4}(m_{4}-1)(m_{4}-2)}N_{5,-4}N_{6,-4}N_{-4,-2}X_{m_{3}+1,m_{4}-3}\\
&& \qquad\qquad + m_{4}m_{5}N_{5,-4}N_{6,-5}X_{m_{5}-1}\\
&& \qquad\qquad +
m_{4}m_{6}N_{5,-4}\left[\frac{1}{2}\sum_{i=1}^{6}(\alpha_{6},-\alpha_{6})(m_{6}-1)+(\alpha_{6},\alpha_{i})n_{i}
\right]X_{m_{4}-1,m_{6}-1}\\
&& \qquad\qquad + m_{4}m_{6}N_{5,-4}\alpha_{6}^{(1)}X_{m_{4}-1,m_{6}-1,k_{1}+1}+
m_{4}m_{6}N_{5,-4}\alpha_{6}^{(2)}X_{m_{4}-1,m_{6}-1,k_{2}+1} \\
&& \qquad\qquad + m_{4}n_{1}N_{5,-4}N_{6,1}X_{m_{4}-1,n_{1}-1,n_{2}+1}\\
&& \qquad\qquad + m_{4}n_{2}N_{5,-4}N_{6,2}X_{m_{4}-1,n_{2}-1,n_{4}+1}\\
&& \qquad\qquad + \frac{1}{2}{m_{4}n_{2}(n_{2}-1)}N_{5,-4}N_{6,2}N_{4,2}X_{m_{4}-1,n_{2}-2,n_{3}+1}\\
&& \qquad\qquad + m_{4}n_{4}N_{5,-4}N_{6,4}X_{m_{4}-1,n_{4}-1,n_{5}+1}\\
&& \qquad\qquad +
m_{5}\left[\frac{1}{2}(\alpha_{5},-\alpha_{5})(m_{5}-1)+(\alpha_{5},-\alpha_{6})m_{6}+\sum_{i=1}^{6}(\alpha_{5},\alpha_{i})n_{i}
\right]X_{m_{5}-1}\\
&& \qquad\qquad +  m_{5}\alpha_{5}^{(1)}X_{m_{5}-1,k_{1}+1}
+  m_{5}\alpha_{5}^{(2)}X_{m_{5}-1,k_{2}+1}\\
&& \qquad\qquad + m_{6}N_{5,-6}X_{m_{6}-1,n_{4}+1}\\
&& \qquad\qquad + \frac{1}{2}{m_{6}(m_{6}-1)}N_{5,-6}N_{4,-6}X_{m_{6}-2,n_{2}+1}\\
&& \qquad\qquad + \frac{1}{6}{m_{6}(m_{6}-1)(m_{6}-2)}N_{5,-6}N_{4,-6}N_{2,-6}X_{m_{6}-3,n_{1}+1} \\
&& \qquad\qquad + m_{6}n_{2}N_{5,-6}N_{4,2}X_{m_{6}-1,n_{2}-1,n_{3}+1}\\
&& \qquad\qquad + n_{1}N_{5,1}X_{n_{1}-1,n_{3}+1} ,\\
&&\rho(E_{6}) X_{(mnk)} = X_{n_{6}+1}\\
&& \qquad\qquad + m_{2}N_{6,-2}X_{m_{1}+1,m_{2}-1}\\
&& \qquad\qquad + m_{4}N_{6,-4}X_{m_{2}+1,m_{4}-1}\\
&& \qquad\qquad + \frac{1}{2}{m_{4}(m_{4}-1)}N_{6,-4}N_{-4,-2}X_{m_{3}+1,m_{4}-2}\\
&& \qquad\qquad + m_{5}N_{6,-5}X_{m_{4}+1,m_{5}-1}\\
&& \qquad\qquad + m_{6}\left[\frac{1}{2}(\alpha_{6},-\alpha_{6})(m_{6}-1)+\sum_{i=1}^{6}(\alpha_{6},\alpha_{i})n_{i}
\right]X_{m_{6}-1}\\
&& \qquad\qquad + m_{6}\alpha_{6}^{(1)}X_{m_{6}-1,k_{1}+1}+ m_{6}\alpha_{6}^{(2)}X_{m_{6}-1,k_{2}+1}\\
&& \qquad\qquad + n_{1}N_{6,1}X_{n_{1}-1,n_{2}+1}\\
&& \qquad\qquad + n_{2}N_{6,2}X_{n_{2}-1,n_{4}+1}\\
&& \qquad\qquad + \frac{1}{2}{n_{2}(n_{2}-1)}N_{6,2}N_{4,2}X_{n_{2}-2,n_{3}+1}\\
&& \qquad\qquad + n_{4}N_{6,4}X_{n_{4}-1,n_{5}+1},\\
&& \rho(H_{1}) X_{(mnk)} =X_{k_{1}+1}+ \sum_{i=1}^{6}\alpha_{i}^{(1)}(n_{i}-m_{i}) X_{(mnk)},\\
&&\rho (H_{2}) X_{(mnk)}
= X_{k_{1}+1}+ \sum_{i=1}^{6}\alpha_{i}^{(2)}(n_{i}-m_{i}) X_{(mnk)}.
\end{eqnarray*}

{

\end{document}